\documentclass[useAMS,usenatbib]{mn2e}

\usepackage{graphicx}
\usepackage{amsmath}
\usepackage{amssymb}
\usepackage{breqn}
\graphicspath{{./figures/}}
\usepackage{verbatim}
\usepackage{enumitem,lipsum}
\usepackage{enumerate}
\usepackage[lofdepth,lotdepth]{subfig}
\usepackage[bottom,hang,splitrule]{footmisc}
\def \Ks {{\textit K_s}} 
\def \sqdeg {\, {\rm deg}^2} 
\def\mic {\, \mu{\rm m}}

\def \appendfigs{\par \vspace*{0.7in}
   \begin{center}\Large{\textbf{APPENDIX: Images}} \end{center}\par  
   \setcounter{figure}{0}
    \renewcommand\thefigure{A1} 
    \def\topfraction{0.05}
    \def\bottomfraction{0.95}
    }  

\bibliography{VikingIds}

\renewcommand{\thefootnote}{\fnsymbol{footnote}}

\title[Herschel-ATLAS: VIKING counterparts]{\textit{Herschel}-ATLAS: VISTA VIKING near-IR counterparts in the Phase 1 GAMA 9h data$^{\star}$}

\author[S. Fleuren et al.]{S. Fleuren$^{1}\dagger$, W. Sutherland$^{1}$, L. Dunne$^{2}$, D.J.B. Smith$^{3}$, S.J. Maddox$^{2}$, \and
J. Gonz\'{a}lez-Nuevo$^{13}$, J. Findlay$^{1}$, R. Auld$^{4}$, M. Baes$^{8}$, N.A. Bond$^{12}$, D.G. Bonfield$^{3}$, \and
N. Bourne$^{28}$, A. Cooray$^{18}$, S. Buttiglione$^{6}$, A. Cava$^{10}$, A. Dariush$^{17}$, G. De Zotti$^{6,13}$, \and
S.P. Driver$^{14,5}$, S. Dye$^{28}$, S. Eales$^{4}$, J. Fritz$^{8}$, M. L. P. Gunawardhana$^{21,22}$,\and
 R. Hopwood$^{17,9}$, E. Ibar$^{19}$, 
R.J. Ivison$^{19,7}$, M.J. Jarvis$^{3,27}$, L. Kelvin$^{5,14}$, A. Lapi$^{13,15}$, \and
J. Liske$^{25}$, M.J. Micha{\l}owski$^{26}$, M. Negrello$^{9,6}$,
 E. Pascale$^{4}$, M. Pohlen$^{4}$, M. Prescott$^{23}$, \and 
 E.E. Rigby$^{26}$,
 A. Robotham$^{5}$, D. Scott$^{24}$, P. Temi$^{11}$,
 M.A. Thompson$^{3}$, E. Valiante$^{4}$, \and
 P. van der Werf$^{16}$  \\
$^{1}$School of Physics and Astronomy, Queen Mary University of London, Mile End Road, London, E1 4NS, UK\\
$^{2}$Department of Physics and Astronomy, University of Canterbury, Private Bag 4800, Christchurch 8140, New Zealand\\
$^{3}$Centre for Astrophysics Research, Science \& Technology Research Institute, University of Hertfordshire, Hatfield, Herts, AL10 9AB, UK\\
$^{4}$School of Physics and Astronomy, Cardiff University, Queens Buildings, The Parade, Cardiff, CF24 3AA, UK\\
$^{5}$SUPA, School of Physics \& Astronomy, University of St Andrews, North Haugh, St Andrews, KY16 9SS, UK\\
$^{6}$INAF –- Osservatorio Astronomico di Padova, Vicolo Osservatorio 5, I-35122, Padova, Italy\\
$^{7}$Institute for Astronomy, University of Edinburgh, Blackford Hill, Edinburgh EH9 3HJ\\ 
$^{8}$Sterrenkundig Observatorium, Universiteit Gent, Krijgslaan 281 S9, B-9000 Gent, Belgium\\
$^{9}$Department of Physics and Astronomy, The Open University, Walton Hall, Milton Keynes, MK7 6AA,
UK\\
$^{10}$Departamento de Astrof\'{\i}sica, Facultad de CC. F\'{\i}sicas, Universidad
Complutense de Madrid, E-28040 Madrid, Spain\\
$^{11}$Astrophysics Branch, NASA Ames Research Center, Mail Stop 245-6, Moffett Field, CA 94035, USA\\
$^{12}$Cosmology Laboratory (Code 665), NASA Goddard Space Flight Center, 
Greenbelt, MD 20771, USA\\
$^{13}$Astrophysics Sector, SISSA/ISAS, Via Bonomea 265, 34136 Trieste, Italy\\
$^{14}$International Centre for Radio Astronomy Research (ICRAR), The University of Western Australia,
35 Stirling Highway, Crawley, \\WA6009, Australia\\
$^{15}$Dipartimento di Fisica, Universita Tor Vergata, Via Ricerca Scientifica 
1, 00133 Roma, Italy\\
$^{16}$Leiden Observatory, P.O. Box 9513, 2300 RA Leiden, The Netherlands\\
$^{17}$Physics Department, Imperial College London, South Kensington Campus, London SW7 2AZ\\
$^{18}$Department of Physics \& Astronomy, University of California, Irvine, CA 92697\\
$^{19}$UK Astronomy Technology Centre, Royal Observatory, Edinburgh, EH9 3HJ, UK\\
$^{20}$Department of Physics and Astronomy, Macquarie University, Sydney, NSW 2109, Australia\\
$^{21}$Australian Astronomical Observatory, PO Box 296, Epping, NSW 1710, Australia\\
$^{22}$Sydney Institute for Astronomy, School of Physics, University of Sydney, NSW 2006, Australia\\
$^{23}$Astrophysics Research Institute, Liverpool John Moores University, Twelve Quays House, Egerton Wharf, Birkenhead, CH41 1LD, UK\\
$^{24}$Department of Physics and Astronomy, 6224 Agricultural Road, University of British Columbia, Vancouver, BC, V6T 1Z1, Canada\\
$^{25}$European Southern Observatory, Karl-Schwarzschild-Str. 2, 85748 Garching, Germany\\
$^{26}$SUPA, Institute for Astronomy, University of Edinburgh, Royal Observatory, Blackford Hill, Edinburgh EH9 3HJ, UK\\
$^{27}$Physics Department, University of the Western Cape, Cape Town, 7535, South Africa\\
$^{28}$Centre for Astronomy and Particle Theory, The School of Physics \& Astronomy, Nottingham University, University Park Campus, \\Nottingham,NG7 1HR, UK}

\begin{document}
\setlength{\parindent}{10pt}
\date{13 March 2012} 

\pagerange{\pageref{firstpage}--\pageref{lastpage}} 
\pubyear{2012v2}

\maketitle

\label{firstpage}

\clearpage

\begin{abstract} We identify near-infrared $\Ks$ band
counterparts to \textit{Herschel}-ATLAS submm sources, 
 using a preliminary object catalogue from the VISTA VIKING survey.
 The submm sources are selected from 
 the H-ATLAS Phase 1 catalogue of the GAMA 9h field,
which includes all objects detected at
250, 350 or $500 \mic$ with the SPIRE instrument. 
We apply and discuss a likelihood ratio (LR)
method for VIKING candidates within a search radius of $10''$ 
 of the 22,000 SPIRE sources with a $5\sigma$ detection at $250\mic$. 
We estimate the fraction of SPIRE sources with a 
 counterpart above the magnitude limit of the VIKING 
survey to be $Q_0 \approx 0.73$. We find that 11,294
($51\%$) of the SPIRE sources have
a best VIKING counterpart with a reliability $R\ge 0.8$, and 
 the false identification rate of these is estimated to be $4.2\%$. We expect to miss $\sim 5\%$ of true VIKING counterparts.
 There is evidence from $Z-J$ and $J-K_s$ colours that the reliable 
 counterparts to SPIRE galaxies are marginally 
 redder than the field population. 
We obtain photometric redshifts for $\sim 68\%$ of 
all (non-stellar) VIKING candidates with a
median redshift of $\tilde{z}=0.405$. We have spectroscopic redshifts for 3147
($\sim 28\%$) of the reliable counterparts from existing redshift surveys.
Comparing to the results of the optical identifications supplied with the
Phase I catalogue, we find that the use of medium-deep near-infrared
data improves the identification rate of reliable counterparts from
 36\% to 51\%. 
\end{abstract}

\begin{keywords}
Methods: Statistical, Submillimetre: Galaxies.
\end{keywords}
\setlength{\footnotemargin}{0.05cm}
\footnotetext[0]{$^{\star}$\hspace{0.08in}\textit{Herschel} is an ESA space observatory with science instruments provided by European-led Principal Investigator consortia and with important participation from NASA. VISTA is an ESO near-infrared telescope in Chile.}
 \footnotetext[0]{$^{\dagger}$\hspace{0.08in}E-mail: s.fleuren@qmul.ac.uk}

\section{Introduction\label{secintro}} 

The extra-galactic universe has been well observed in the optical and, 
 to an extent, the near-infrared wavelength ranges for several decades; 
 in contrast, only relatively recently have we started to 
 carry out unbiased surveys in the sub-millimetre (submm) 
  wavelengths \citep{sma97,mor05,dev09} 
 and we still have a limited understanding of the sources 
  responsible for the bulk of submm emissions. 
Finding optical/near-infrared counterparts enables us to  
  complement our knowledge with optical/near-infrared colours, 
 photometric redshifts, and potentially follow up 
 with multi-object spectroscopy. \par
Using near-infrared instead of optical wavelengths to identify submm 
 sources allows us to probe galaxies out to higher redshifts 
($z \ge 0.5$) where rest-frame visible bands are shifted to
 the observed near-infrared.  This is especially important for the 
 dusty galaxies expected to be detected by instruments onboard 
 the Herschel Space Observatory (HSO, \citealp{pil10}). 
The dust in those galaxies absorbs most UV photons and re-radiates 
 in the far-infrared and submm wavelengths. \par
In the near-infrared restframe, the $K$ band ($\sim 2 - 2.3 \mic$) samples 
the peak of the emission of the older stars and is hence well suited to 
evaluate the stellar mass of a galaxy \citep{col01,bel03}. 
The well established link between stellar mass and specific star formation 
 rate (SSFR, \citealp[][for recent evidence with \textit{Herschel}]{rod10}) 
ensures that the $K$ band is also interesting when investigating the SFR.\par
 One difficulty in combining object catalogues at widely different
 wavelengths  lies in deciding which objects
 are truly associated, and which are unrelated 
  foreground/background objects. 
Matching submm sources to objects observed in much shorter wavelengths
 is particularly difficult, because the large submm beam sizes, 
and hence the comparatively lower angular resolution, 
and high confusion noise increase the positional uncertainties, 
which in turn forces us to increase the radius we employ searching 
 for plausible counterparts. 
The increased search radius, together with the high surface density 
of objects from optical/near-infrared surveys, 
 is responsible for the ineffectiveness of a simple closest 
neighbour method. \par
A method often applied to previous submm surveys
 consists of first matching submm sources 
to radio or mid-infrared sources \citep{ivi07,dye09,big11}, 
before utilising the multi-wavelength data that are already 
 available for the radio/mid-infrared counterparts, 
 or using the more accurate positions to improve on the matching technique.
 This is advantageous
because of the radio/far-infrared correlation \citep{hel85,hai10,ivi10,jar10,mic10,bou11}, 
 the lower surface density 
 and the high positional accuracy of radio 
 catalogues  \citep{ivi07,dye09,dun10}.
 Unfortunately, identifying optical/near-infrared counterparts 
  via radio counterparts is not yet practical for the 
 Herschel Astrophysical Terahertz Large Area Survey (H-ATLAS, \citealp{eal10}) 
 since current radio telescopes cannot deliver the required 
 area/depth combination; \citet{har10} found 187 radio sources 
  within the H-ATLAS SDP (Science Demonstration Phase) field, 
 less than 3\% of the H-ATLAS sources.
 Radio surveys will dramatically improve in future 
 with SKA (Square Kilometre Array) and 
  its precursors LOFAR (LOw Frequency ARray, http://www.lofar-uk.org), 
 ASKAP (Australian Square Kilometre Array Pathfinder, \citealp{deb09}), and 
 MeerKAT (\citealp{sch08,joh08}), but not for several years. \par
 In the mid-infrared, \citet{ros10} have used Spitzer source positions 
 for the source extraction process and deblending 
 of HerMES (Herschel Multi-Tiered Extragalactic Survey, \citealp{oli10}) sources, 
 taking advantage of the small positional uncertainties at 
 3.6$\mic$ and 24$\mic$. However, 
 mid-infrared data are available for only small patches 
 within the H-ATLAS observed fields, see \citet{bonprep}, 
 using WISE (Wide-field Infrared Survey Explorer, \citealp{wri10}) 
 data and \citet{kimprep}, using {\it Spitzer}-IRAC data.
\par 
An alternative approach, adopted here, is to match the submm sources 
 directly to a near-infrared catalogue by using information on 
 the positional uncertainty probability distribution of the
submm sources and the magnitude distribution of the optical/near-infrared
 objects utilising a likelihood ratio method (LR). 
This approach uses the ratio
of the probabilities of a match being the true counterpart
 and being an unrelated background object
\citep{deR77,pp83,wol86,ss92}. We describe the likelihood ratio method in
detail in section~\ref{secLR}.\par 
The paper is structured as follows. Section
2 describes the surveys and the data selection. The LR method is discussed
in detail in section 3, and in section 4 we explain the method to obtain
our photometric redshifts. 
Section 5 presents our identification results, and in section 6 we
compare our results with those of the optical matching supplied with
 the SPIRE Phase 1 catalogue. 
Section 7 summarises our conclusions. 
 
\section{data\label{secData}}

\subsection{\textit{Herschel}-ATLAS SPIRE sources\label{secdataSpire}}
The Herschel-ATLAS survey \citep{eal10} 
 is a large open-time key project carried out with the Herschel Space 
 Observatory. The full survey will cover approximately $550 \sqdeg$ of
 high galactic latitude sky in six patches; 
 the survey covers the wavelength range $100- 500\mic$, 
 providing imaging and photometry. 
 Two instruments survey in 5 passbands, centred on wavelengths 
 100 and $160 \mic$ (PACS, \citealp{pog10}) 
 and 250, 350 and $500 \mic$ (SPIRE, \citealp{gri10}). 
 The beams have FWHM (full width at half maximum) of respectively 8.7'', 13.1'', 18.1'', 25.2'' and 36.9'',
 with 5$\sigma$ point source sensitivities of 132, 126, 32, 36 and 45 mJy
  in the above 5 passbands.
 The maps and data reduction are discussed in 
 detail in \citet{pas10} and \citet{iba10}, and the source 
 catalogue creation is described in \citet{rig10}.\par
 The H-ATLAS fields have been selected for low cirrus foreground, and
 overlap with a number of other  
  existing and planned surveys to profit from multi-wavelength data.
 A few important overlapping surveys are the 
 Sloan Digital Sky Survey (SDSS, \citealp{yor00}) in the optical, 
 the Galaxy And Mass Assembly (GAMA, \citealp{dri11}) survey which includes
 a spectroscopic redshift survey of mostly SDSS objects 
 \citep[see][for the target selection]{bal10} and the 
 VISTA VIKING imaging survey \citep{sutprep} in the near-infrared. \par
Here we use the H-ATLAS Phase 1 catalogue of the GAMA 9h (G09) 
 field ($\approx$ 54 deg$^2$), comprised of 26,269 sources 
detected at $5\sigma$ in the 250$\mu m$ band. 
The H-ATLAS catalogue supplies optical counterparts 
from SDSS \citep{hoyinPrep} using a similar LR technique to that 
presented here, as discussed in \citet{smi10}. 
We find that 22,000 of these SPIRE positions are within the region 
 observed by VIKING up to late 2010 
(an area of approximately $50 \sqdeg$, contained within RA between 
 128 and 141 degrees and Dec between $-2$ and $+3$ degrees), and these   
comprise our catalogue used in the matching hereafter. 

\renewcommand{\thefootnote}{\arabic{footnote}}\setcounter{footnote}{0}

\subsection{VISTA VIKING data \label{secDataVista}}
VISTA is a 4m wide-field telescope at the ESO Paranal 
 observatory in Chile \citep{eme10}. 
The camera has 16 near-infrared detectors and an instantaneous field of 
 view of $0.6 \sqdeg$, and its filter set includes
 the five broad-band filters \textit{Z},\textit{Y},\textit{J},\textit{H},$\Ks$ with 
  central wavelengths $0.88 - 2.15 \mic$.
The VISTA Kilo-degree INfrared Galaxy (VIKING, Sutherland et al., in prep.) 
 survey is one of the public, large-scale surveys ongoing with VISTA. 
 The survey has been planned to cover around $1500 \sqdeg$ of 
 extragalactic sky in the above five filters, including 
 one southern stripe (including the H-ATLAS SGP stripes), 
  one equatorial strip in the North galactic cap 
 (including the GAMA 12h and 15h fields) and also the GAMA 9h field. 
 The median image quality is $\approx 0.9''$, 
 and typical $5\sigma$ magnitude limits are
  $J \approx 21.0, \Ks \approx 19.2$ on the Vega system or 
 $J \approx 21.9, \Ks \approx 21.0$ on the AB system. \par
The data processing \citep{lew10} is a collaboration of the Cambridge
  Astronomy Survey Unit (CASU) and the Wide Field Astronomy Unit
  (WFAU) at the Royal Observatory in Edinburgh. 
 The data used in this paper is from the Vista Science Archive (VSA) 
 produced and maintained in Edinburgh, released internally 
 on the 14th April 2011; this is our preliminary object catalogue. 
 The VSA builds on the WFCAM Science Archive (WSA), providing
  similar access and functionality\footnote{for a 
 detailed description of the functionality and the access options, 
 see \citet{ham08} } (e.g. image cut-outs, SQL queries etc.).
  Sources are extracted after the merging of individual frames and are listed
  in tables together with astrometric and photometric information.\par
For our object catalogue, we require a $5\sigma \ \Ks$ band detection, 
 using (aperture corrected) aperture photometry with a diameter of $2''$.
   We also require a $J$ band detection to exclude the large majority of
  spurious detections (bright star halos, satellite trails etc). 
 In addition, we only use sources that are primary detections 
 (best source in overlap regions) with error flags smaller
  than 256 (only informational error quality conditions, e.g. deblended), and sources flagged as saturated or noise were excluded. While the above constraints on the object catalogue are necessary, given its size, to obtain a fairly clean sample, inevitably, we will lose some objects that could be true counterparts. The majority of those lost would be around bright stars, and we expect this fraction to be around 2\%.
 This data selection results in 1,376,606 objects in our VIKING catalogue 
 from the G09 field.

\subsection{Star-galaxy separation\label{secsgsep}} 
The
likelihood ratio method we employ to match both catalogues (see section 3),
uses the magnitude distribution of the true counterparts which will depend
 on the morphological type of the VIKING objects. 
 The VSA uses a shape parameter calculated
from the brightness profile of the objects and the point-spread-function (PSF)
on each individual detector to classify objects as stars and galaxies. The
galaxy sample is optimised for completeness, leading to a stellar sample that
is optimised for reliability. 
 We have hence started building our stellar
 sample by using the VSA stellar probability $\ge 95\%$. This puts
464,033 objects firmly into the stellar class. \par
\begin{figure}
\centering
\includegraphics[width=0.5\textwidth]{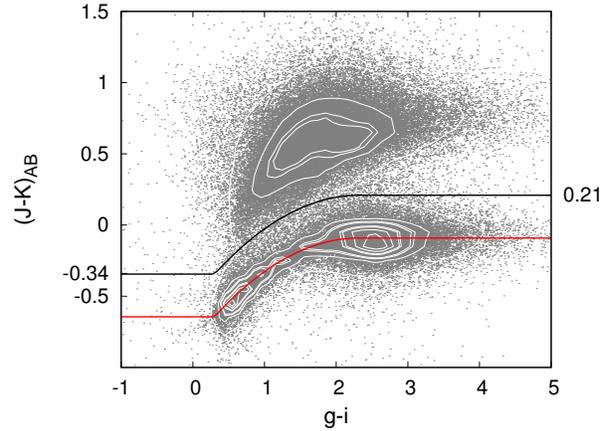}
\caption{--- The colour-colour diagram of VIKING objects with SDSS
counterparts. The red line shows the stellar locus, obtained from
fitting a quadratic equation in the range [0.3,2.3] to the sample with
VSA probability of being a star $>0.95$. The black line is offset +0.3
 mag from the locus, 
 representing the adopted star/galaxy separation cut.  Objects
 without SDSS identifications are classified according to their $J-Ks$ colour
 only, leaving a stripe of $-0.34 < (J-\Ks)_{AB}  < 0.21$ where 
  the colour-based classification is ambiguous. }
\label{figCCD}
\end{figure}

The remaining objects were then matched to the SDSS database, using
  the nearest neighbour within $2''$, to obtain $g-i$ colours for a
  classification on the $J - \Ks$ vs $g-i$ colour-colour diagram.
This follows the procedure used by   
 \citet{bal10} to select a highly complete
 galaxy sample for the GAMA input catalogue,
in conjunction with a SDSS shape parameter. For objects with $r<19.8$,
we employ their prescription for the classification, using a combination
of shape and colour parameters. However, most of our objects are
  much fainter than the $r<19.8$ cut used by \citet{bal10}, and here
 the SDSS morphological classification is unreliable,   
 as was evident when we used a sub-sample
 with available spectroscopic redshifts. For objects with an 
 SDSS counterpart fainter than 
  $r>19.8$, we then classify via the position on
the colour-colour diagram. Fig.~\ref{figCCD}
shows the colours of the VIKING sample, the location of our stellar locus
 and the chosen separation line. \par 

 The remainder of VIKING objects that have not been classified above, 
 i.e. objects with VSA stellar probability $<0.95$ and 
 without an SDSS counterpart, are separated as follows: objects with 
 $ (J - \Ks)_{AB} >0.21$
are classified as galaxies and those with $(J- \Ks)_{AB} < -0.34$ 
 as stars. The logic of
this can be seen on the colour-colour diagram: even without $g-i$ information,
 these objects must lie respectively above/below the black separation
 line in Fig.~\ref{figCCD}.   This leaves a stripe
  at intermediate $J - \Ks$ colour where the colour 
  classification remains ambiguous:  
 just over 12,000 objects fall into this category. 
 Here, we look at the VSA shape
 classification again and relax our earlier cut of 95\% to 70\%,
 classifying objects as stars or galaxies with a cut at $pStar = 0.7$.  
 Finally, we move
  573 objects classified as stars to the galaxy class as they have confirmed
  non-stellar redshift of $z>0.002$ from SDSS spectra:  those are mostly 
  confirmed
 QSOs. This classification results in a sample of 847,530 galaxies 
  and 529,076 stars\footnote{In the
catalogue, we include the flag `sgmode' which indicates how we have arrived
at the classification: \textbf{1} VSA star with pStar$>0.95$, \textbf{2}
uses \citet{bal10} for objects with $r<19.8$, with slight modifications,
\textbf{3} colour-colour selection for objects with $r\ge 19.8$, 
 \textbf{4}
 $(J - \Ks)_{(AB)}$ colour selection for objects without SDSS counterpart within
 $2''$, \textbf{5 / 6} 
 VSA pStar$ = 0.7$ cut for objects without SDSS counterparts
  and ambiguous in $J -\Ks$ colour.  
 The sgmode flag is changed by appending a zero to
the initial flag if the object was moved from the star class to the galaxy
class having a confirmed non-stellar redshift.}.\par Using this method of
classification, QSOs without spectra in SDSS are mainly classified as stars,
selected by morphology. This is clearly not ideal, because objects in the star
sample are less likely to be identified as reliable counterparts to the SPIRE
sources. A more detailed separation, taking QSO properties into account, will
be explored in future work \citep{hoyinPrep}. 
\section[]{Applying the Likelihood Ratio method \label{secMethod}}
\subsection[]{The Likelihood Ratio method \label{secLR}}
One of the earliest approaches \citep{deR77,pp83,wol86} to the matching of two source catalogues in different wavelengths uses the ratio of two
likelihoods: the likelihood that a true counterpart is observed at a distance
$r$ from the source and with magnitude $m$, 
  and the likelihood that an unrelated
background object is observed with the same properties: 
\begin{equation}
 L = \frac{P(r,m,id)}{P(r,m,chance)}
\end{equation} 
 The probability $P(id|r,m)$ that an object at distance $r$ and
with magnitude $m$ is a true counterpart, also called the reliability, is then:
\begin{equation} \label{eqOldRel}
 P(id|r,m) = \frac{L}{L+1}
\end{equation} using Bayes' theorem and the theorem of total probability.
\par
\cite{ss92} extended the likelihood ratio by  
 incorporating information about other potential counterparts to one
 source in the calculation of the reliability, and 
also including additional information (denoted by $c$) which can be, 
 for example, colour information or star/galaxy classification.
This is especially useful
in a situation where the matching catalogue has a high surface density
so that there is a high probability of there being more than one possible
counterpart. The reliability $R_j$ that the $j$th candidate for one
  source is the true
counterpart is then: \begin{equation}
 \label{eqNewRel}
R_j=P(id|r,m,c)_j = \frac{L_j}{\sum_i L_i + (1 - Q)} 
\end{equation} 
where
 $i$ runs over all candidates for this source, and $Q$ is the probability
 (for a random source) of finding a genuine counterpart above the 
  limiting magnitude of the matching survey. 
In contrast to equation~\eqref{eqOldRel}, this includes information about
other candidate counterparts to the source.
 To clarify this distinction, 
  equation~\ref{eqOldRel} is the reliability {\em without}
 information about other candidates for the same source (e.g. picking
  one candidate at random from a concatenated list 
  of all candidates for a large number of submm sources),
  while equation~\ref{eqNewRel} is the reliability given the set of 
  all candidate matches for one given source. \par
 \citet{ss92} define $q(m,c)$,
the probability distribution of the true counterparts with magnitude $m$
and additional property $c$, and $f(x,y)$ the probability distribution of
 the source positional errors $x,y$, 
 normalised such that 
\begin{equation}
 \label{eqDefQ}
  \int \int_{-\infty}^{+\infty} f(x,y)\, dx \, dy = 1 \hspace{0.05in}
\text{and}\hspace{0.02in}
 \int_{-\infty}^{M_{lim}}q(m,c)\, dm\, dc = Q \end{equation}
where $M_{lim}$ is the limiting magnitude in the matching catalogue. 
 It is usually assumed that the positional errors are
  independent of the magnitude and other additional information, so that
$P(m,x,y,c,id)=P(m,c,id) P(x,y,id)$.  If $n(m,c)$ is the surface
 density of unrelated background objects per unit magnitude, 
  the likelihood ratio for any candidate match is then:
\begin{equation}
 \label{eqLR}
 L=\frac{P(m,c,x,y,id)}{P(m,c,x,y,chance)}=\frac{q(m,c)f(x,y)}{n(m,c)}
\end{equation} 
 In practice, the probabilities above have to be
 estimated from the data by fitting simple models. 
 The surface density of unrelated
  background objects is estimated from the surface density of objects in the
 matching catalogue.
  The next two subsections explain how we estimate 
 the distributions $f(x,y)$ for the positional errors and $q(m)$ 
 for the true counterparts.
 
\subsection{Positional uncertainties\label{secPos}} 
We here adopt the simple model that the H-ATLAS 
 source positional errors are Gaussian with
 equal RMS $\sigma_{pos}$ in each of RA and Dec; 
  then the normalisation condition above requires  
\begin{equation} 
 \label{eqf}
 f(x,y) = f(r) =\frac{1}{2\pi\sigma_{pos}^2}
  \exp(-\frac{r^2}{2\sigma_{pos}^2}) \ . 
\end{equation} 
  where $r = \sqrt{x^2+y^2}$ is the radial position difference , 
  and we note $f(r)$ has units of (solid angle)$^{-1}$.  
 \citet{smi10} have estimated the positional errors of $> 5\sigma$
SPIRE sources, using histograms in RA and Dec of the total number of SDSS
sources within a 50'' box around the SPIRE 250 $\mu$m centres and taken
the clustering of the SDSS objects into account\footnote{\citet{smi10}
fit the sum of the Gaussian positional errors and the clustering signal,
convolved with the Gaussian errors, to the resulting histograms. For a more
detailed description of the derivation of the positional uncertainties, see
section 2.1 of their paper. }. To be able to use their results, we measure
the correlation of our VIKING objects to the SDSS objects, constructing
the corresponding RA and Dec histograms, 
 with VIKING objects within a box
around the SDSS positions. 
 The 1$\sigma$ VIKING position errors are $<0.2''$ and
therefore negligible compared to the SPIRE errors.
  We hence adopt the weighted
mean 1$\sigma$ positional uncertainty of $\sigma_{pos}=2.40''\pm0.09$
quoted in \citet{smi10} and assume the errors to be symmetric in RA
and Dec.\par In theory, the positional uncertainty should depend on the SNR (signal-to-noise ratio) and the FWHM of
the observations. \citet{ivi07} derive\footnote{a derivation of this formula
can be found in the appendix of \citet{ivi07}} the positional uncertainty as
$\sigma=0.6\times \frac{\rm{FWHM}}{\rm{SNR}}$. Following	\citet{smi10}, we adjust the
 formula to match our mean positional error by inserting a scaling factor of
$1.09$, so that: 
\begin{equation} 
\label{eqSigPos}
  \sigma_{pos}=0.655 \times \frac{\rm{FWHM}}{\rm{SNR}}
\end{equation} 
with the SPIRE mean FWHM$=18.1''$. For each SPIRE source, the positional uncertainty from equation~\ref{eqSigPos} is then used in our LR calculation. We also set a minimum of
 $\sigma_{pos} = 1''$ for sources with high SNR, 
  as there are limitations to the minimum positional accuracy
 from SPIRE and SDSS maps, as discussed in \citet{smi10}.
  We adopt a conservative search radius of $10''$ which would include 
 $99.996\%$ of the real counterparts assuming Gaussian errors; 
  in practice, there is evidence for slightly non-Gaussian 
 wings \citep[see][]{hoyinPrep}, but
  this radius still includes almost all genuine matches. 

\subsection{Estimation of $q(m,c)$ and $Q$ \label{secQ}} 
We estimate the
probability distribution $q(m,c)$ of the true counterparts by using the
background subtracted sample of candidate matches, as outlined in \citet{cil03}. For
each class ($c=$ galaxies and $c=$ stars), we estimate $q(m)$ from the data
as follows:
 \begin{list}{$\bullet$}{\setlength\labelwidth{0.2in}\setlength\leftmargin{0.2in}} 
 \item create 
 a magnitude distribution $total(m)$ of all objects within a search
radius of 10'' around the SPIRE sources 
\item background subtract $total(m)$
to obtain the so-called $real(m)$ distribution. 
\item normalise $real(m)$
so that $q(m)=\frac{real(m)}{\sum_{m_i} real(m_i)}\times Q_0$ 
\end{list} 
where we sum over bins of magnitude.The background is determined from the
  number density $n(m)$ measured from the whole catalogue, scaled to a 10'' circle.  
 The normalisation factor $Q_0$ is an estimate of the 
 probability of finding a
 counterpart in the VIKING survey down to the $5\sigma$ survey limit, $Q$ in
 equation~\eqref{eqDefQ}\footnote{\citet{cil03} have introduced the constant $Q_0$ as the value of $Q$ as estimated from the data.}. 

A reasonably accurate value of $Q_0$ is important, since this enters
 the reliability formula above, equation~\ref{eqNewRel}. 
 Simply estimating $Q_0$ via
 a stacking analysis (summing $real(m)$ and dividing by the 
 number of SPIRE sources) is not ideal, since source clustering and/or
 genuine multiple counterparts will tend to overestimate $Q_0$ by 
 multi-counting, and
  therefore reliability estimates will be biased high. \par
To avoid this multi-counting problem, 
 we decide to estimate $1-Q_0$, the fraction of SPIRE sources 
 without a VIKING-detected counterpart, hereafter called blanks:  
 these will be mostly real sources fainter than the VIKING limit, 
  but also including counterparts outside the search radius and
  spurious SPIRE detections (if any). 
 We start by counting the observed blanks to a given search radius $r$; 
  we then need to correct for those sources that have VIKING candidate matches,
  where the match(es) are in fact unrelated to the SPIRE source. 
 The number of true blanks is the number of observed blanks, plus 
  the number of true blanks that have been matched with a random VIKING
  object.
  To estimate the latter, we create a catalogue of
 N (=number of SPIRE sources) random positions and cross-match 
 with the VIKING catalogue. 
Hence, defining $\bar{R}$ as the number of blanks 
at random positions, 
 $R$ the number of random positions with a VIKING source within 10" and $\bar{S}$ the number of 
 observed SPIRE blanks,
we can calculate the number $\bar{S}_t$ of true SPIRE blanks as follows: 

\begin{equation}
\bar{S}_t=\bar{S} + \big[ \bar{S}_t\times \frac{R}{N} \big]\Leftrightarrow
\bar{S}_t=\frac{\bar{S}}{1-R/N}=\frac{\bar{S}}{\bar{R}/N} \end{equation}

Dividing by N provides us then with the fraction of the SPIRE sources that
are true blanks, $\bar{S}_t / N =\bar{S}/\bar{R}$, which is our estimate for
 $1-Q_0$. Thus, we only need to divide the number of SPIRE blanks 
 by the number of random blanks, for a given search radius. \par
\begin{figure}
\centering \includegraphics[width=0.5\textwidth]{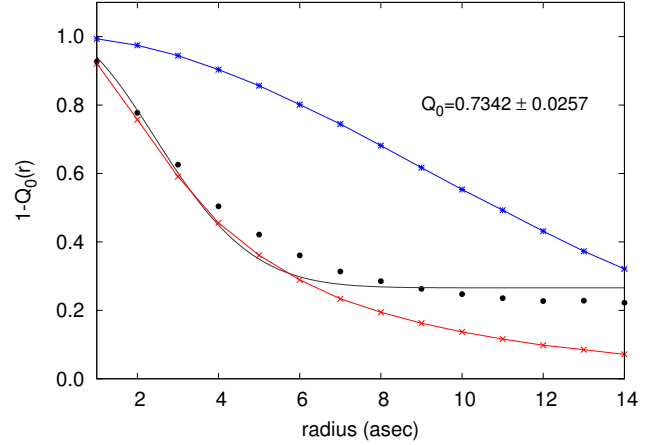} 
\caption{---
 Illustration of the procedure to estimate $1-Q_0$. The red crosses show the
fraction of blank SPIRE sources within search radius $r$, while
the blue stars show the same fraction for random positions.  The black
filled circles represent the data points obtained from dividing the number
of blank SPIRE positions by the number of blank random positions,
our estimate for $(1 - Q_0)F(r)$. The black line represents the best
fit to the model $(1 - Q_0)F(r)$, see 
  equation~\eqref{eqDefQ}, with $Q_0=0.7342\pm
0.0257$.} \label{figQest} 
\end{figure}

For our default search radius of 10'', we obtain $1-Q_0=0.25$, or $Q_0=0.75$. 
We could use this value in our subsequent LR analysis; 
 however, it depends on the 
value of the search radius and the $Q$, as defined in
equation~\eqref{eqDefQ}, of the VIKING catalogue is independent of the search
radius. 
 We would like to obtain an estimate $Q_0$ of $Q$ that is independent of 
 the radius and so
 repeat the above procedure for radii in the range 1-15''. The values we
 obtain for the fraction of true blanks are shown in Fig.~\ref{figQest} as 
 black points. \par
 We then
model the dependence of the true blanks on the search radius as follows: 
 a SPIRE
 blank at radius $r$ is a source whose counterpart is
 either fainter than the VIKING limit, or lies outside
the search radius, or both. The former probability is $1-Q_0$, this
is the first term in equation~\eqref{eqQModel}. The probability of the
counterpart to reside outside the search radius can be calculated from
the positional error distribution $f(r)$, leading to our second term in
equation~\eqref{eqQModel}. The third term follows if we assume 
 that both possibilities are independent of each other,
 and using the standard probability result 
  $P(A\ {\rm or}\ B) = P(A) + P(B) - P(A\ {\rm and}\ B)$.   
 Our model for the dependence of the true blanks
 on the search radius $r$ is then: 
 \begin{equation}
 \label{eqQModel}
\begin{split} &(1-Q_0)+(1-F(r))-(1-Q_0)(1-F(r))=1-Q_0F(r)\hspace{1
in}\\ & \text{where} \hspace{0.1in}F(r)=\int_0^r
P(r')dr'=1-e^{-\frac{r^2}{2\sigma^2}}\\ & \text{and} \hspace{0.1in}P(r)=2\pi
rf(r) \end{split}
 \end{equation}  
 Fitting this model to the data, we obtain
$Q_0=0.73\pm 0.03$ as our best-fitting value. Fig.~\ref{figQest} shows the
best fit model as the black line and the black filled circles as our
data points ($\bar{S}/\bar{R}$ for each radius). \par
The model underestimates
the number of SPIRE blanks in the data in the range $4'' \leq
 r \leq 8''$ and overestimates the data for $r > 10''$. 
This might show some evidence for clustering of the VIKING objects
which we have not explicitly considered, but which is accounted for 
 in the value of the mean positional 
 uncertainty $\sigma_{pos}=2.4''$ by \citet{smi10}; they convolve their
  model with the clustering signal of the SDSS sources to obtain the value we adopt here. 
It might also
demonstrate that the Gaussian approximation for the SPIRE positional errors,
our equation~\eqref{eqf}, is not entirely accurate. 
 This is also evident
when investigating histograms of distances 
 of VIKING and SDSS objects to SPIRE sources.
  We see slightly higher numbers of objects at distances of
 around 10'' than expected if we assume a Gaussian error distribution. This
assumption is examined and will be discussed in \citet{hoyinPrep}.\par

 Our fitted value of $Q_0=0.73$ is consistent with the value of $Q_0=0.75$
from the datapoint at our search radius of $10''$, and is more conservative.
  We hence adopt the fitted value $Q_0=0.73$ for our subsequent LR analysis.
\par
Having estimated the general value of $Q_0$, we still need to know the individual contributions from stars and galaxies, $Q_{0s}$ and $Q_{0g}$ respectively, with $Q_0=Q_{0s}+Q_{0g}$.
A drawback of the above approach is that we cannot separate our blank SPIRE fields into stars
and galaxies. Therefore, we estimate the value of $Q_{0s}$ in applying equation (10) of \citet{smi10}, using a background subtracted sample of possible matches within 10'',
yielding a value of 0.004.
This is very small indeed and shows how unlikely
it is that stars are detected with SPIRE. For our LR analysis we then
adopt $Q_{0g}=0.72$ and $Q_{0s}=0.01$, the values for $Q_0$ used in the
 normalisation of $q(m)$ for galaxies and stars respectively.  
\subsection{Probability of mis-identifying a
true counterpart 
\label{secWrong}} 
Given the above model, we now estimate $P(wrongID|m_{true})$,
 defined to be the probability that a true counterpart
  with a given VIKING $\Ks$ magnitude $m_{true}$
is not the best candidate using our LR method. 
 This situation occurs, if
 the true counterpart has a likelihood ratio value $L=L_i$ and there
 exists a chance match with $L>L_i$ for the same source. 
 Hence: \begin{equation} \label{eqPropWrong}
\begin{split} P(wrong ID|\, m_{true})=&\int_0^{L_{max}}P(L_i|m_{true})\\
&\times P(chance >L_i)\, dL_i \end{split} \end{equation}

The probability $P(chance > L_i)$ that there exists a chance match
 with $LR > L_i$ for one SPIRE source 
 can be estimated through simulations. The steps of the procedure we have
used are as follows: 
 \begin{list}{$\bullet$}{\setlength\labelwidth{0.2in}\setlength\leftmargin{0.2in}} 
\item Create $N$ random positions on
an area common to both VIKING and H-ATLAS in the G09 field
\item For each random position, calculate the likelihood ratio $L$ for
random matches, if any, in the VIKING catalogue.  
\item Create the distribution $D(L)$
 of the highest likelihood ratio value for each random position.  
\end{list}
 From the probability 
distribution $D(L)$ of the highest LR values for candidate matches to random positions,
we can calculate the probability that a given source
 has a chance candidate above $L_i$ by chance: \begin{equation}
P(chance > L_i )=\frac{1}{N}\int_{L_i}^{L_{max}}D(L)\, dL \end{equation}

\begin{figure} \centering
\includegraphics[width=0.5\textwidth]{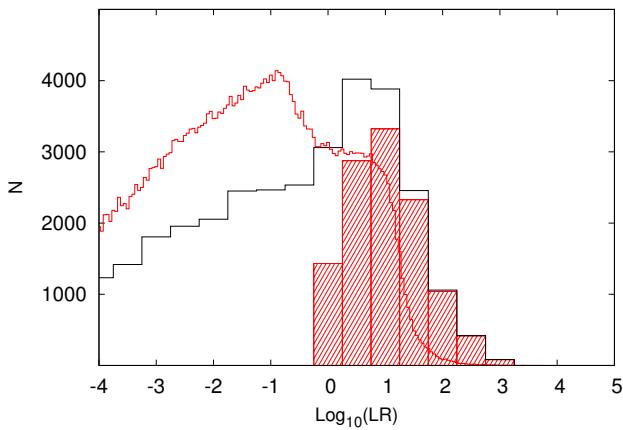} \caption{
\label{figDLR}--- The solid red line shows the distribution $D(L)$ of
the highest $L$ values of candidate matches to $>10^6$ random positions in the G09
field. The peaks at lower and higher $L$ values are due to star and galaxy
matches respectively. The black histogram represents the LR values of all
candidate matches to the SPIRE positions. The filled red histogram represents the
reliable matches only. } \end{figure}

A similar method has been employed by \citet{dye09} to calculate
the probability of radio associations to the BLAST (Balloon-borne Large Aperture Submillimetre Telescope)  submm sources.
Fig.~\ref{figDLR} shows the resulting distribution $D(L)$
from $>10^6$ random positions, together with the histogram of
the likelihood ratio values of the candidate matches to the real SPIRE positions and of
the subset of reliable counterparts (e.g. objects with a probability of $>80\%$ of being the true counterpart, see section~\ref{secResultsKs}).\par 
The probability $P(L_i|m_{true})$,
the first factor in the integral in equation~\eqref{eqPropWrong}, that a true
counterpart with $\Ks$ magnitude $m_{true}$ acquires the likelihood ratio $L_i$
can be calculated analytically from the probability distribution $P_r(r)=2\pi
r f(r)$ of the offsets. For a given $m=m_{true}$ we have $f_i=f(r)=L_i\times
n(m_{true})/q(m_{true})$ and hence $P(L_i|m_{true})=P_f(f_i)$. We can find
the probability distribution $P_f(f)$ from $P_r(r)$ by performing a variable
transformation: 
\begin{equation} P_f(f)=-P_r(r(f))\times \frac{dr}{df}=2\pi
\sigma^2 \end{equation}

This is a surprisingly simple result, i.e. for Gaussian errors, 
 $P_f(f)$ has a uniform distribution between 0 and its maximum value
  $1/2\pi \sigma^2$. 
We would like to note though that
the positional uncertainty depends on the value of the SNR. For
simplicity, we have used the value of $\sigma_{pos}=2.4''$, allowing for
most of the SPIRE sources to have a SNR value close to 5. A more detailed
analysis could take into account the probability distribution of the SNR
values of all SPIRE sources.\par 
 Having calculated the probability of a
true counterpart with assumed magnitude $m_{true}$ not being the best candidate in our LR method, we can then calculate $P(wrongID)$, 
 the same probability
 integrated over the model magnitude distribution of the true counterparts: 
 \begin{equation}
 P(wrongID)=\frac{\int_0^{M_{lim}} P(wrongID|m)\, q(m)\, 
  dm}{\int_0^{M_{lim}}q(m) \, dm}
\end{equation}
  Using the distribution $D(L)$ from the random positions,
our VIKING sample as described in section~\ref{secDataVista} to derive the
magnitude distributions $q(m)$ and $n(m)$, as well as $\sigma_{pos}=2.4''$,
we obtain $P(wrongID)=0.0493$. 
We hence expect to mis-identify around $\sim 5\%$ of the true 
 VIKING counterparts to the SPIRE sources. 

This is likely to
overestimate the true value because we have not taken 
 the individual SNR values
of the SPIRE sources into account. This performance measure of the LR method is
compared in section~\ref{secResultsKs} with the 
 false identification rate for reliable counterparts.

\section[]{Photometric redshifts \label{secPhot}}
The submm wavebands benefit from negative k-correction with the effect
that objects can be detected at least out to
$z \gtrsim 5$ in the $870\mic$ band  \citep{blr93,cha11}. 
 Selecting sources at $250\mic$ reduces the effectiveness
 of the negative k-correction, but models still predict a significant
  fraction of the SPIRE sources to reside 
 at $z > 1$ \citep[e.g.][]{amb10,lap11}. \par
A significant fraction of our VIKING counterparts ($\sim 28\%$)
 have spectroscopic
redshift from the GAMA and SDSS redshift surveys. 
 For the remainder, we
will obtain photometric redshifts by combining the near-infrared photometry
with optical photometric information from the SDSS survey. 
 Objects with spectroscopic redshift from
the GAMA survey are then employed as a training set to estimate
  the photometric redshifts using neural networks, see below.

SDSS matches are
found for $\sim 68\%$ of the VIKING candidates within $2''$ by performing a simple nearest
neighbour match. The remainder of the VIKING objects are too faint to be detected
 in the SDSS survey. For near-complete visible detections 
we have to wait for the 
VST KIDS survey\footnote{PI Konrad Kuiken at Leiden University}, which will
observe the VIKING areas down to 24.8, 25.4, 25.2 and 24.2 (10$\sigma$, AB)
in ${u', g', r', i'}$ bands respectively, thus giving detections in at least
  $r', i'$ bands for nearly all VIKING objects.  
\par 
With a search radius of $2''$ we are
unlikely to identify a wrong SDSS counterpart. Using the surface density
of the SDSS catalogue employed for the optical matching by \citet{smi10},
we estimate around 2\% (170 objects) of our subsample of reliable VIKING
matches to have a wrong SDSS identification within $2''$.\par 
For photometric redshifts, we use a neural network
  method in which the photometry of
 objects with available spectroscopic redshifts provides the sample
 to train a network. 
Once trained, the network is then used to obtain photometric redshifts from
 the photometry of objects without reliable redshift information. 
The photometry in the different bands employed can differ, for
  instance we are using SDSS modelmags and VIKING Vega aperture 
magnitudes. 
In contrast, a template fitting method where 
 the photometry is compared to the expected photometry from a class of different (empirical and/or theoretical) SEDs (Spectral Energy Distribution), needs to use very carefully calibrated photometry in all bands. The software
used to obtain the photometric redshifts is ANNz \citep{col04}, a publicly available product. We train a committee of 3 networks for each possible photometric band combination, recommended to minimise the network variance. The output of ANNz is the photometric redshift for each object together
with a redshift error estimate which takes the photometric errors in each
band into account.\par
We use 30,697 objects with photometry from SDSS and VIKING and 
 spectroscopic redshifts from GAMA
in our training set. 
 The median spectroscopic redshift of the GAMA sample 
 is $\tilde{z}_{spec}=0.211$. The median redshift of the 
 VIKING counterparts is expected to be higher and the training set should reflect this.
 Currently, there are no deeper spectroscopic surveys available that
 overlap with the VIKING area. 
 There are though some deeper spectroscopic redshifts from surveys that are within the UKIDSS LAS
\footnote{UKIDSS LAS is carried out with the UKIRT WFCAM, 
and images an area of $\sim 2500 \sqdeg$ in 
 the \textit{YJHK} filters to a depth $K$=18.4} area \citep{law07,hew06}. 
 Therefore, we undertake a
comparison of WFCAM and VISTA photometry, to be able to use deeper photometric
information from the zCOSMOS\footnote{zCOSMOS is a redshift survey carried out
on the ESO VLT with the VIMOS spectrograph on the COSMOS field} \citep{lil07}
and DEEP2\footnote{DEEP2 is a redshift survey carried out with the Keck
telescopes with a pre-selected redshift range of $0.75-1.4$.} \citep{dav03}
surveys. We match VIKING objects in a 4 deg$^2$ area in the H-ATLAS SDP field
to objects in the UKIDSS LAS
 and obtain the mean and standard error for the difference in magnitudes
(apermag3) in each of the bands \textit{YJHK}. 
We are then able to use the LAS photometry for these samples 
 by subtracting this mean from the LAS magnitudes and adding the standard error of the difference in quadrature to the photometric error. This
allows us to use the deeper photometry and spectroscopic 
 redshifts of zCosmos
(1530 objects) and DEEP2 (238 objects), which overlap with the UKIDSS LAS
survey (but not with VIKING), as a training set.\par 
Adding all our subsamples
together, we obtain an overall training catalogue with 32,465 spectroscopic
redshifts. We then compile a photometric catalogue for all VIKING candidate matches
with \textit{ugriz} photometry (modelmags) and \textit{YJHK} photometry
(aperture magnitudes), where present. We attempt photometric redshifts
where we have at least 2 infrared bands with good photometry.\par 
Using the deeper spectroscopic redshifts
from zCOSMOS and DEEP2 forces us to exclude the VIKING \textit{Z} band in
the training catalogue. Investigating how this influences our photometric
redshifts, we create a second training catalogue, this time using just the
GAMA subset (95\% of the first training catalogue), so that we can include
the VIKING \textit{Z} band. The scatter in the differences 
 between photometric and spectroscopic redshifts for objects with a spectroscopic redshift is slightly lower for this second training set, but this is to be expected because we compare only the lower redshift end. 
 Assuming that the inclusion of deeper redshifts into the training set reflects the true VIKING redshift distribution better, we adopt the photometric redshifts from our first training catalogue, the GAMA-zCOSMOS-DEEP2 training set.\par

 \section{Results\label{secResults}} 
\subsection{$\Ks$ band matching  \label{secResultsKs}}
There are 22,000 SPIRE sources within the sky area corresponding to the G09
 VIKING object catalogue. Of those, 18,989 sources have at least one
possible match within a search radius of $10''$ in the VIKING $\Ks$ 
 band selected catalogue, with a total of 35,800 candidate matches;
 of which 30,659 are classified as galaxies (85.6\%)
 and 5,141 are classified as stars (14.4\%), as described 
 in section~\ref{secsgsep}. 
Table~\ref{tabNoMatches} shows the number of SPIRE sources matched 
 as a function of the number of
candidate matches found per position.\par
\begin{figure} \centering
\includegraphics[width=0.5\textwidth]{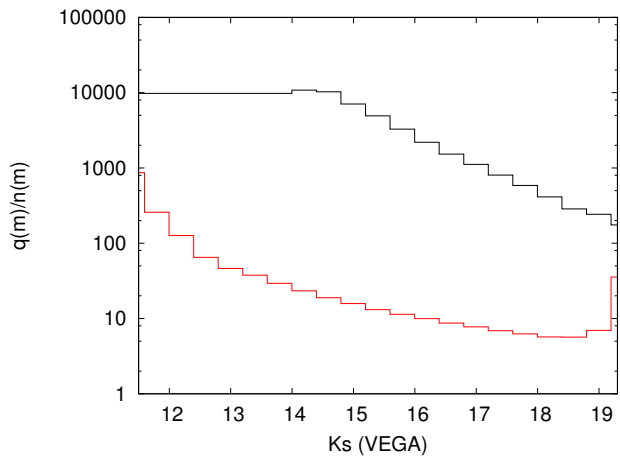} \caption{--- The magnitude
dependent part $q(m)/n(m)$ of the LR analysis, calculated from the data as
described in section~\ref{secQ}. The black line indicates the galaxy candidates,
the red line shows the stellar candidates. The values for the bright bins for the
galaxy distribution were extrapolated from the first bin that included more
than 10 galaxies, at $m=13.4$.} \label{figQn} \end{figure}

\begin{figure} \centering \includegraphics[width=0.5\textwidth]{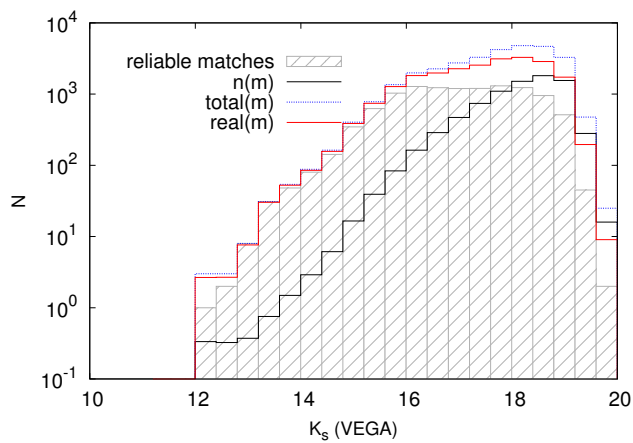}
\caption{--- The magnitude distributions involved in estimating $q(m)$
for galaxies. Here $n(m)$ (solid black) is the distribution of the background
objects, calculated from the whole VIKING G09 catalogue, as described in
section~\ref{secsgsep}, and $total(m)$ (dashed blue) 
 is the magnitude distribution
of all possible matches within $10''$. 
 $real(m)$ (solid red) is the background
subtracted distribution as described in section ~\ref{secQ} and is significantly brighter than the background. The light
grey shaded histogram represents the magnitude distribution of the reliable
matches. } \label{figBtrGals} \end{figure}

\begin{figure} \centering \includegraphics[width=0.5\textwidth]{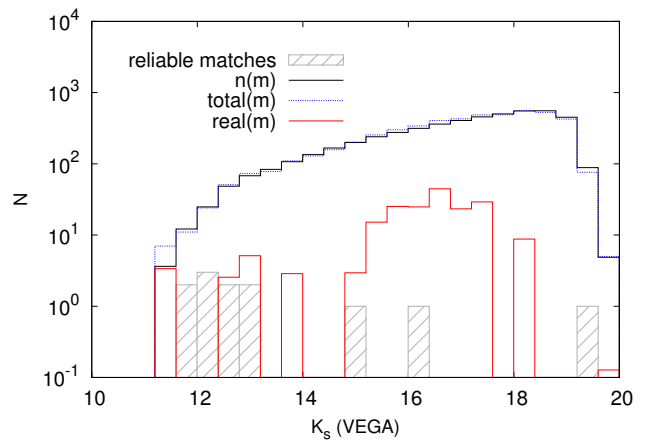}
\caption{--- The magnitude distributions involved in estimating $q(m)$ for
stars. The colour scheme is as in Fig.~\ref{figBtrGals}.} \label{figBtrStars}
\end{figure}

 
There are 
 11,294 SPIRE sources with a best VIKING counterpart with a 
 reliability $>80\%$ (11,282 galaxies/12 stars). 
 This means we were able to match $\sim 51\%$ of the SPIRE sources
 with a high reliability. We will refer to the set of matches with $R>0.8$ as
 ``reliable'' hereafter 
 (as we show later, the mean reliability of this set 
   is near 0.96).   
Fig.~\ref{figQn} shows the magnitude dependent distribution
$q(m)/n(m)$ for galaxies and stars used in the calculation of the LR
values. Fig.~\ref{figBtrGals} and ~\ref{figBtrStars} show the magnitude
distributions involved in estimating
$q(m)$: of the background $n(m)$, of the possible matches
$total(m)$ and of the background subtracted sample $real(m)$. Also shown in the figures is the magnitude distribution for reliable
counterparts. The reliable matches show a lower fraction of fainter
counterparts compared to all candidate matches, representing the steep increase
of fainter objects in the background number counts, causing
  lower reliability values  for fainter matches. 
 \par\par 

We can estimate a false ID rate by summing up the
 complement of reliability values of the reliable matches: 
 \begin{equation}
 N_{falseID}=\sum_i (1-R_i)=469.25 
\end{equation} 
corresponding to a mean reliability of $0.958$
 and a false ID rate of $4.2\%$. \par 
In the appendix, we show cut-outs of VIKING and SDSS images around
SPIRE sources for 9 positions drawn at random from our reliable sample.
\begin{table}
\centering
 \caption{The distribution of the
 number of VIKING $\Ks$ band sources within 10'' of the 250 $\mu m$
 SPIRE positions. Of the 22,000 sources in the VIKING area, 8118
 have only one possible match within 10'' and $59.7\%$ of these
 are determined to be reliable. This emphasises the difference from a simple
 nearest neighbour match.} \label{tabNoMatches} \small{ \begin{tabular}{crrrr}
  \hline
   N(matches)	  &  N(SPIRE)  & N(reliable)  &  \%  \\
 \hline 0 & 3011 &  &  \\
1 & 8118 & 4851 & 59.76\\ 2 & 6619 & 4040 & 61.04\\ 3 & 2968 & 1710 & 57.61\\
4 & 968 & 529 & 54.65\\ 5 & 241 & 128 & 53.11\\ 6 & 63 & 31 & 49.21\\ 7 &
11 & 5 & 45.45\\ 8 & 1 & 0 & 00.00\\ \hline
 Totals & 22,000 & 11,294 & &\\ \hline
\end{tabular}} 
\end{table}
\subsection[]{VIKING and SPIRE colours\label{secColours}}
\begin{figure}
\centering
\includegraphics[width=0.5\textwidth]{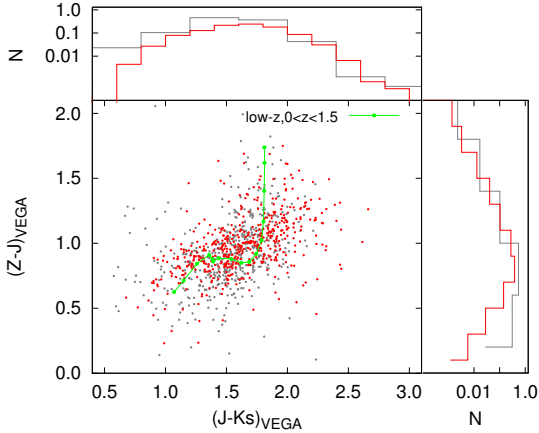}
\caption{--- VIKING colour-colour diagram of reliable counterparts (red) and background galaxies (grey) with 5$\sigma$ detections in $Z$,$J$ and $\Ks$. The horizontal and vertical histograms show the distributions in $J-\Ks$ and $Z-J$ respectively. The green points connected by a green line represent the evolutionary track of a typical H-ATLAS galaxy SED for $z<0.35$ from Smith et al. (2011b) and calculated by \citep{fin12}.}
\label{figZJK}
\end{figure}

 Fig.~\ref{figZJK} shows the \textit{ZJK} colour-colour diagram of the 10,121 reliable counterparts (red) with 5$\sigma$ detection in all 3 VIKING bands. Colours from randomly selected background galaxies are depicted in grey. 
The redshift evolution of the submm selected mean galaxy template of 
\citet{smiinPrep} is shown in green. The template has been artificially 
redshifted between $z=0-1.5$ in intervals of $dz=0.1$ and colours have been 
computed by integrating the product of the template SED with the VISTA 
response functions at each redshift interval. A small deviation from the 
Vega system is present in the VISTA $Z$-band and a measured offset 
\citep{fin12} has been added to the colours computed here to 
reflect this.\par 

 The median $Z-J$ and $J-\Ks$ colours for the reliable matches are $0.97$ and $1.64$; for the background objects the median colours are $0.89$ and $1.54$. Performing a two-sided K-S test on the $J-\Ks$ and the $Z-J$ colours for reliable matches and the background objects enables us to reject at a significance of 99.96\% that the two populations are drawn from the same distribution for either colour. Thus, we find evidence that the reliable matches to the SPIRE sources are slightly, but significantly redder than the population of all VIKING objects in the G09 field.\par
\begin{figure}
\centering
\includegraphics[width=0.5\textwidth]{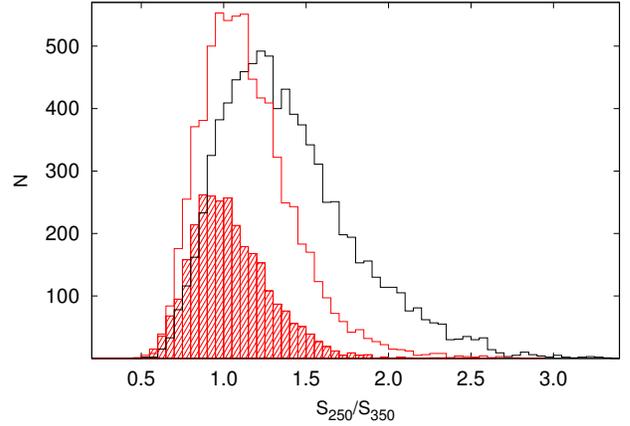}
\caption{--- Distribution of the SPIRE $S_{250}/S_{350}$ colour. Black: sources with reliable matches. Red: sources that have unreliable matches only. Red shaded: blank sources. The blank sources show distinctively redder colours than the other two populations, suggesting they are at higher redshifts.}
\label{fig250o350}
\end{figure}

\begin{figure}
\centering
\includegraphics[width=0.5\textwidth]{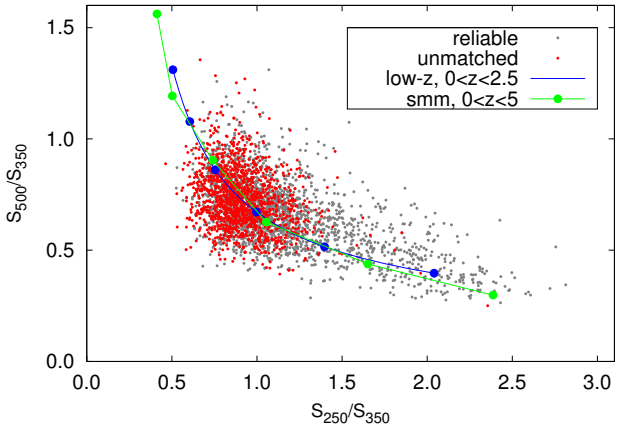}
\caption{--- Colour-colour diagram of SPIRE objects with $5\sigma$ detection in $250\mu m$ and $350\mu m$ and $3\sigma$ detection in $500 \mu m$. The dark grey points represent SPIRE sources that have reliable counterparts, the red points show the colours of blank SPIRE sources. The blue points connected by the blue line represent the colours of a typical H-ATLAS galaxy SED for $z<0.35$ from \citet{smiinPrep}. The dots are at an interval of $z=0.5$, starting at $z=0$ on the right hand side. The green points connected by the green line represent the colours of the submm SED in \citet{lap11}. The points are at an interval of $z=1$, starting with $z=0$ on the right.}
\label{figRatio}
\end{figure}

Within the search radius of $10''$, 3011 sources have 
 no 5$\sigma$ $\Ks$ VIKING candidate, i.e.  they are 
 almost certainly fainter than the VIKING limit. 
 Fig.~\ref{fig250o350} shows $S_{250}/S_{350}$ histograms 
 separately for the sources with a reliable match (best $R > 0.8$),
   sources with non-reliable match(es) (best $R < 0.8$), 
 and for sources that are 
 blank in VIKING (black, red and shaded red respectively). 
 The blank sources (median=1.01) have distinctively 
 redder colours than the sources with reliable 
 matches (median=1.32), suggesting that they reside at 
 higher redshifts with the peak of the dust
  emission moving to longer wavelengths. 
 The colours of sources with unreliable matches 
 (median=1.11) lie in between the other two 
 populations, indicating that they might be composed 
 from members of both populations.\footnote{The SPIRE fluxes for Fig.~\ref{fig250o350} (and Fig.~\ref{figRatio}) have not been corrected for confusion or Eddington boosting. Both effects are negligible for the 250$\mic$ band, but become more pronounced in the 350 and 500$\mic$ bands. If we do correct the fluxes, the median values for the ratios in Fig.~\ref{fig250o350} are shifted by +0.1; this does not affect our conclusions. A slight shift towards higher values is also seen in Fig.~\ref{figRatio} and again, this does not affect any of the results.}
 
 From our value of $Q_0$, we 
 expect around 60\% of the unreliably matched sources to have a 
 true counterpart, but for which we do not have a high enough 
 reliability, with the remaining 40\% being matched to unrelated 
 background objects. \citet{smi10} find a similar trend 
 in the distribution of the $S_{250}/S_{350}$ colour of  
 SPIRE sources in the SDP field matched to the 
 SDSS $r$-band catalogue for the 3 different populations.\par 
 We also show a SPIRE colour-colour diagram of the sources 
 in Fig.~\ref{figRatio}. 
 The colours of the  reliably matched 
 and the blank sources are very similar to those in \citet{smi10}, their fig. 9. This figure can also be compared to fig. 1 in \citet{amb10}. We add the evolutionary tracks of two templates: a low-z template compiled by \citet{smiinPrep} from optical counterparts to SPIRE sources out to $z\simeq 0.35$ (blue line) and the submm template from \citet{lap11} (green line), thought to be appropriate for high-z H-ATLAS sources at $z>1.2$.
From the tracks of the two SED templates we can again suggest 
that the blank sources lie at higher redshifts in general than the sources with reliable matches.
 Assuming that the H-ATLAS sources are comprised of two distinct populations, see section~\ref{secResultsPhot}, a lower redshift population with mainly normal galaxies (with a much higher star-formation rate than local normal galaxies), and a higher redshift population of dusty submm galaxies (likely to be giant proto-spheroidal galaxies in the process of forming most of their stars), our blank sources could represent a mixture of the former at $z\gtrsim 0.7$ and the latter at $z\gtrsim 1.5$.\par
\subsection[]{Multiple counterparts\label{secResultsMultiples}}
The LR method assumes that there is only one true counterpart to each 
 source, and assigns reliabilities self-consistently based on this, so that the
 sum of reliabilities cannot exceed 1. 
 Thus, if more than one counterpart with $R > 0.2$ is present, 
 we will not find a reliable match. 
 If individual reliabilities add up to our threshold of $0.8$, 
 we could assume that these candidate matches are all associated with the 
 sources, either through confusion or in a real physical 
 sense (i.e. merging galaxies). \par
In our results, we find 1444 SPIRE sources that fulfill 
 the above criteria and potentially have multiple true counterparts 
 in VIKING. Most of those will have additional close chance objects within 
 the search radius, denying a reliable identification, 
 but some might be genuine mergers or constitute members of the same 
 cluster. 
 We can rule out a chance match by comparing the redshifts 
 of all possible matches to one SPIRE source. Checking for available 
 redshifts in the GAMA and SDSS spectroscopic redshift databases, 
 we find matches to 37 sources whose redshifts are 
 within 5$\%$ of each other. The mean redshift difference 
 is 0.0011 with a maximum difference of $\bigtriangleup z=0.0187$. 
 Those could be either merging galaxies or members of the same cluster.\par
\begin{figure}
\centering
\includegraphics[trim = 90mm 7mm 100mm 7mm, clip, width=0.5\textwidth]{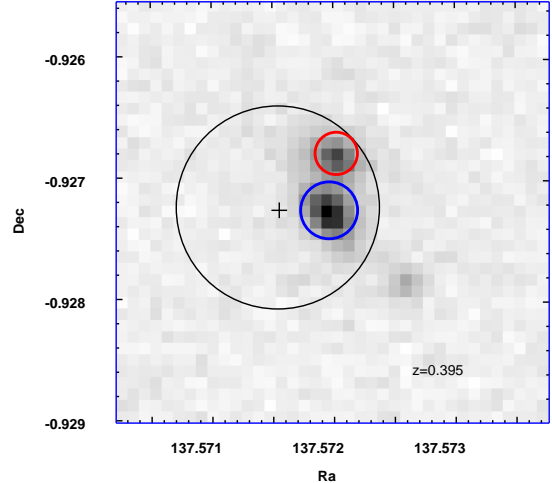}
\caption{ --- VIKING $\Ks$ image of HATLAS J091017.1-005538, 12'' on the side. The black cross indicates the SPIRE position and the black circle represents the 2$\sigma=2.9''$ positional uncertainty of this SPIRE source. The blue circled object has reliability $R=0.74$, the red circled object has $R=0.25$. The third candidate has negligible reliability, but could still be part of the interacting system. }
\label{figInteracting}
\end{figure}

We use our photometric redshifts, see section~\ref{secResultsPhot}, to select further candidates. We account for the higher errors in the photometric redshifts by allowing a redshift difference of 10$\%$ and also compare photometric redshifts of possible matches for which we do not have a spectroscopic redshift with spectroscopic redshifts of other candidates to the same source. We find 602 further sources where the SPIRE flux potentially originates from an interacting system or from galaxies within the same cluster. Fig.~\ref{figInteracting} shows a VIKING $\Ks$
  image cutout around one of the those sources, HATLAS J091017.1-005538. Due to the uncertainties in the photometric redshifts, those sources can only be regarded as candidates and need further investigation to be confirmed as interacting systems or as members of the same cluster.\par
It would be interesting to confirm how many sources definitely do not have physically related multiple counterparts and are just unreliable matches, but this is difficult due to the sparsity of available spectroscopic redshifts and the uncertainties on the photometric redshifts. However, we can estimate the number of reliable identifications we are missing due to potentially multiple counterparts. From Table~\ref{tabNoMatches}, we can see that the identification rate for reliable counterparts is approximately 60$\%$ without the presence of additional potential matches. We do not see a decrease in the identification rate for sources with two possible matches, suggesting that the true number of merging galaxy pairs is indeed low. For sources with higher numbers of possible matches, we have an increased possibility of having observed a galaxy cluster and so the identification rate for reliable matches falls. For instance, sources with 4 possible matches have an identification rate of around 55\%, suggesting that we miss 5\% of the reliable matches, equivalent to 50 sources. Adding up the missed reliable matches of all sources with more than 2 possible counterparts, suggests that we are missing around 150 reliable VIKING counterparts due to additional matches within our search radius. This is a small number indeed, only around $10\%$ of the number on our list of candidates for true multiple counterparts. There is evidence, from observation and simulation, for the merger rate to evolve with redshift and to peak at $z\simeq 1.2$ (\citealp{bel06,lot07,rya08}), beyond the redshift of most of our reliable counterparts. Hence, we might miss a substantial fraction of mergers not because we find multiple candidates but rather because they hide in the fraction of blank SPIRE sources. This implies that our candidate list comprises mostly either chance alignments of galaxies or cores of clusters of galaxies resulting in confusion when observed with SPIRE. 


\subsection[]{Stellar matches}
\begin{figure}
\centering
\includegraphics[width=0.5\textwidth]{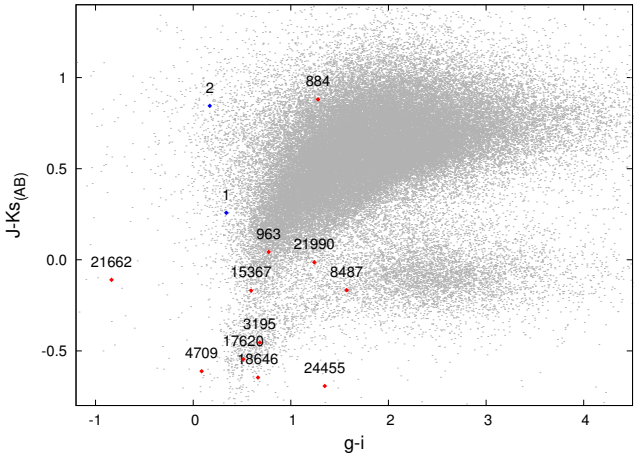}
\caption{ --- Reliable stellar matches, represented by the red dots. For clarity we have labeled the dots with row numbers from the SPIRE catalogue and not with the full HATLAS names. HATLAS J090450.5-014525 (884) has galaxy-like colours and was originally selected to be a QSO target of the Sloan spectroscopic survey but has subsequently been rejected as a target. It is listed as a QSO in the Quasar and Active Galactic Nuclei catalogue by \citet{ver10} with redshift $z=1.005$. The colours of the other objects are consistent with being star-like. The extreme
  $J- \Ks =-2.67$ value of HATLAS J091233.9-004549 is not shown in the diagram and could be due to saturation.
The blue dots represent the colours of the VIKING counterparts to the two blazars found in the SDP field, $1=$H-ATLAS J090910.1+012135, $2=$H-ATLAS J090940.3+020000}
\label{figStars}
\end{figure}

The far-infrared/submm mainly detects cold, dusty objects.
 It is unlikely that stars are detected with SPIRE, unless they are post-AGB, shrouded in dust or have debris disks \citep[e.g.][]{tho10}. We have matched 12 SPIRE sources reliably to point-like objects. Their location on the $J-\Ks$
  vs $g-i$ colour-colour diagram is displayed in Fig.~\ref{figStars}.\par
HATLAS J090450.4-014525 (884 in Fig.~\ref{figStars}) displays galaxy-like colours and is listed as a QSO in the quasar catalogue of \citet{ver10}. All other objects are consistent with having star-like colours. \par
Two blazars were identified by \citet{gon10} in the SDP field from cross-matching to radio observations. We match H-ATLAS J090910.1+012135 reliably to a point-like VIKING object (sgmode$=10$, i.e. point-like object classified as a galaxy on the basis of a non-stellar spectroscopic redshift). 
H-ATLAS J090940.3+020000 is not matched reliably within our search radius of 10''. We find one possible VIKING counterpart within 10'' of this SPIRE source: a bright point-like object ($\Ks=15.03$, sgmode$=10$), lying at a distance of nearly 9'' from the SPIRE position and therefore obtaining a low reliability. However, it lies within 1'' of the known blazar PKS 0907 +022, the object identified as the blazar counterpart to H-ATLAS J090940.3+020000 by \citet{gon10}. Despite the low reliability, there is hence evidence that our VIKING object is the counterpart to H-ATLAS J090940.3+020000.
The colours of both VIKING objects are shown in Fig.~\ref{figStars} as blue dots. Blazars in the H-ATLAS Phase 1 fields are currently investigated by a team led by Marcos Lopez-Caniego with 14 candidates identified so far.\par
For the brighter stars, it is a possibility that the measured SPIRE flux originates from a galaxy that is too faint in the near-infrared to be detected by VIKING, or obliterated by the star, and the star is a chance projection.
 The reliabilities of star counterparts
 are on average lower ($\bar{R}=0.85$) than for the reliable galaxy matches, due to the lower $Q_0$ value and the lower values of $q(m)$ (see Fig.~\ref{figQn}).\par
\subsection[]{Photometric redshift distribution\label{secResultsPhot}}
\begin{figure}
\centering
\includegraphics[width=0.5\textwidth]{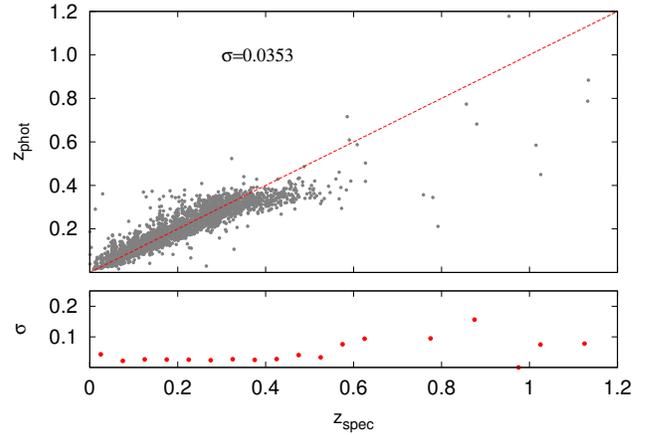}
\caption{ --- Upper panel: photometric vs. spectroscopic redshift for 3147 reliable VIKING counterparts with spectroscopic redshift from either GAMA or SDSS. Lower panel: scatter of the difference $|z_{spec}-z_{phot}|$ for the same 3147 objects in redshift bins of $\delta z=0.05$. }
\label{figPhotozSpecz} 
\end{figure}
We have measured the photometric redshifts for all of our 11,294 reliable VIKING counterparts, 8750 of which have SDSS matches within 2'', as described in section 4. Spectroscopic redshifts exist for 3147 of the reliable VIKING counterparts which allows us to evaluate the accuracy of the photometric redshifts, see Fig.~\ref{figPhotozSpecz}. The spectroscopic redshifts are taken from either the GAMA or the SDSS redshift survey. Where a redshift exists in both surveys for a VIKING object, we have used the GAMA redshift (quality flag $Q\ge3$ only). Excluding 31 confirmed QSOs with $z>1$, the scatter of the difference between our photometric redshifts and the spectroscopic redshifts is $\sigma = 0.0353$. This reduces to $\sigma = 0.0259$ for the normalised redshift distribution $\bigtriangleup z/(1+z_{spec})$. \par
Also visible from Fig.~\ref{figPhotozSpecz} is the tendency for the photometric redshift to underestimate the redshift at higher values of $z$; for $z>0.3$, the bias amounts to $-0.0889$. The systematic underestimation is also found when comparing spectroscopic redshifts with the photometric redshifts from the H-ATLAS Phase 1 catalogue which used optical and near-infrared photometry from SDSS and UKIDSS-LAS, see \citet{smi10}. They have employed a similar training set which suggest that we face an issue with the representativeness of our training set. The reason for the bias seems less likely to be a lack of spectroscopic redshifts at $z>0.3$ but could rather be related to a difference in the colour distribution of galaxies in the training set and our VIKING galaxies. Clearly, more work is needed to investigate the reasons for the bias at $z>0.3$ and, crucially, to assemble a more representative training set which is outside the scope of this paper.\par 
The redshift differences $z_{phot}-z_{spec}$ of the 31 confirmed QSOs with $z>1$ are considerably worse, as the training set includes few high-z QSOs
 and also the near power-law spectra of QSOs means that QSO photo-z estimates
 are much worse than for galaxies.

\begin{figure}
\centering
\includegraphics[width=0.5\textwidth]{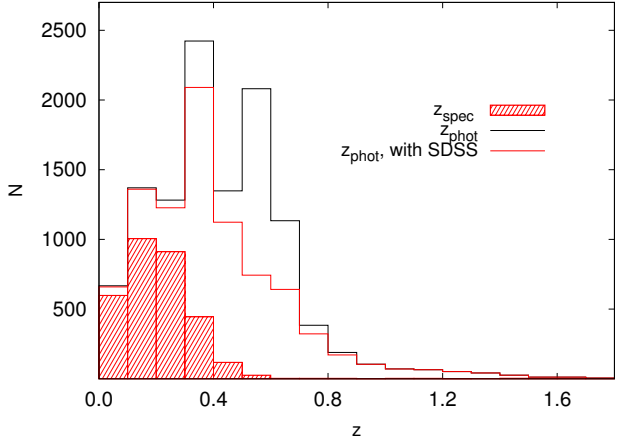}
\caption{--- Redshift distributions for the reliable VIKING matches. The filled boxes represent the spectroscopic redshifts for 3147 of the 11,294 reliable VIKING matches with either GAMA or SDSS spectroscopic redshift with a median redshift of $\tilde{z}=0.199$. The excess peak of all photo-z at $z\sim 0.5$ (black histogram) is formed of redshifts to VIKING objects without optical photometry. } 
 \label{figRedDistr}
\end{figure}

 Currently, we cannot estimate the accuracy of the photo-z of VIKING counterparts with just near-infrared photometry due to the lack of spectroscopic redshifts. In general, they have higher photo-z, as can be seen from Fig.~\ref{figRedDistr}. 
  \par
 
\begin{figure}
\centering
\includegraphics[width=0.5\textwidth]{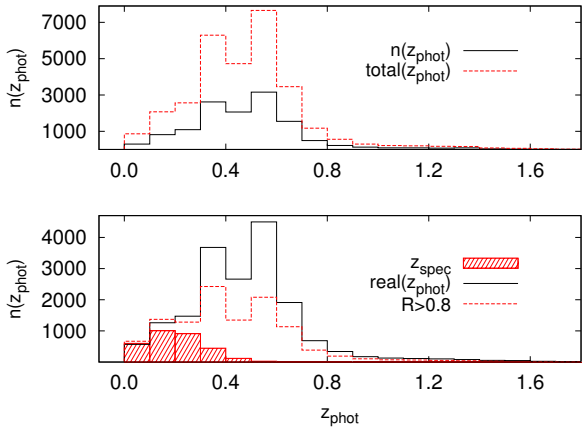}
\caption{--- Upper panel: the dashed red histogram shows the photometric redshift distribution of VIKING background objects, the black line represents the photometric redshift distribution of all VIKING objects within our search radius ($total(z_{phot})$). Lower panel: the expected photometric redshift distribution of the true VIKING counterparts ($real(z_{phot}$, black line), calculated by subtracting the background from $total(z_{phot}$. The filled histogram represents spectroscopic reshifts and the red dashed histogram is the distribution for our reliable counterparts. } 
 \label{figRedCompl}
\end{figure}
Certainly, we are missing more reliable identifications at the higher redshift end than at lower redshifts. The lower panel in Fig.~\ref{figRedCompl} shows a comparison of the expected redshift distribution, $real(z_{phot})$ of the VIKING counterparts to the redshift distribution of our reliable counterparts ($R>0.8$). The expected distribution $real(z_{phot})$ can be calculated in a similar way as the magnitude distributions: from the $total(z_{phot})$, the photometric redshift distribution of the VIKING objects within our search radius of $10''$ and the background distribution $n(z_{phot})$, both shown in the upper panel of Fig.~\ref{figRedCompl}. Table~\ref{tabCompleteness} shows the fraction of reliable to expected counterparts per redshift bin, i.e. the completeness of our photometric redshift sample.\par
\begin{figure}
\centering
\includegraphics[width=0.5\textwidth]{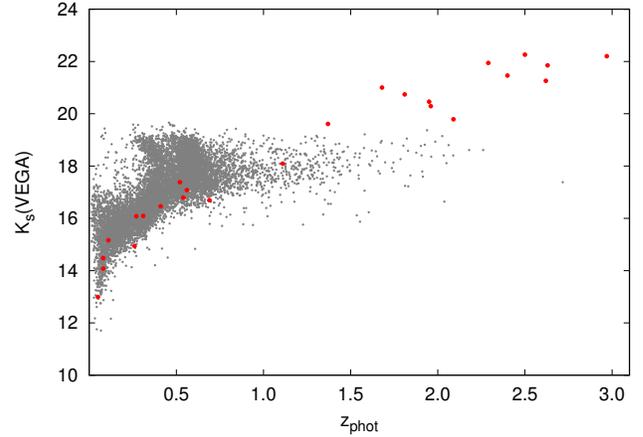}
\caption{--- $\Ks$ magnitudes (Vega) versus photometric redshift. Grey dots: our reliable VIKING counterparts. Red points: BLAST sources of \citet{dun10}. This suggests that we are fairly complete out to $z\sim 1$.} 
 \label{figBLASTzKs}
\end{figure}
\citet{dun10} have calculated very accurate photometric redshifts for counterparts to 20 bright BLAST sources, using a wide range of deep multi-wavelength data in the GOODS-North field. BLAST used a proto-type SPIRE camera and the BLAST sources were selected down to $36 \rm{mJy}$ in 250 $\mic$. We compare our reliable sample to their data on a $\Ks$ vs $z_{phot}$ in Fig.~\ref{figBLASTzKs}. This plot suggests that we are fairly complete out to redshift $z\sim 1$ and are missing higher redshift counterparts due to the VIKING survey limit.\par
\begin{table}
\centering
 \caption{ Photometric redshift completeness, calculated from a background subtracted sample of possible counterparts, $real(z_{phot})$. The errors assume that Poisson errors in the $real$ distribution are dominant.} \label{tabCompleteness} \small{ \begin{tabular}{ccc}
  \hline
   $z_{phot}$	  &  Completeness($\%$)  & $\sigma_{comp}$  \\
 \hline 
0.0-0.1 & 117.8 & 5.8\\
0.1-0.2 & 108.6 & 3.8\\
0.2-0.3 & 87 & 2.9\\
0.3-0.4 & 65.9 & 1.4\\
0.4-0.5 & 50.6 & 1.3\\
0.5-0.6 & 46.3 & 0.9\\
0.6-0.7 & 59.3 & 1.8\\
0.7-0.8 & 56.1 & 2.7\\
0.8-0.9 & 55.8 & 3.6\\
0.9-1.0 & 61 & 5.6\\
1.0-1.1 & 55 & 5.6\\
\hline
\end{tabular}} 
\end{table}
Work carried out by \citet{amb10} using SPIRE and PACS colours, suggests that the SPIRE source redshift distribution might be bimodal, formed by a low redshift population of spirals and a high redshift population of starburst galaxies peaking at $z\sim 2$. \citet{dar11} found that most of the H-ATLAS low-z galaxies are comprised of blue/star-forming galaxies with some highly dusty, red spirals. Evidence for this bimodality in the redshift distribution can also be found in \citet{mad10} using the angular correlation function of 250, 350 and 500 $\mu m$ selected SPIRE sources, and from theoretical models \citep{lag03,neg07}. \citet{smi10} discuss this in more detail in their section 3.3 on the photometric redshift distribution.\par 
In this work, we find that the large majority of our candidate identifications
   have $z_{phot} < 1$,  as expected from the $\Ks$ magnitude limit of the VIKING data.
From our value of $Q_0$, we expect about 27\% of the SPIRE sources to be too faint to be detected in VIKING and hence very likely to be at higher redshifts, see section~\ref{secColours}. Taking into account the possible underestimation of our photometric redshifts at $z>0.3$, we find that $\sim 2\%$ of the H-ATLAS 250 $\mic$ sources with a reliable counterpart lie at $z>1$. At least a similar fraction of sources without a reliable match and expected to have a true counterpart in VIKING, should obtain $z>1$. We hence expect $\gtrsim 30\%$ of our H-ATLAS sources to be found at $z>1$. We compare this with the redshift distributions found for BLAST sources and for sources detected by SPIRE in the GOODS-North field. \par
\citet{dun10} and \citet{cha11}, using 250 $\mic$ BLAST sources, both find that $\sim 50\%$ of their sources lie at $z>1$ from a variety of photometric redshifts, even though the shapes of the distributions differ, with \citet{dun10} seeing a more pronounced bi-modality and \citet{cha11} a greater tail beyond $z=2$. This comprises a significantly higher fraction of high redshift sources than in this work. \citet{dye09} find $\sim 30\%$ of their BLAST sources within a deep field to be at $z>1$, fully consistent with our fraction. Having used different selection criteria for their BLAST sources, either signal-to-noise cut-offs or flux limits in any of the 250, 350 or 500 $\mic$ BLAST bands, leads to different sub-samples that are difficult to compare. In addition, the methods to obtain photometric redshifts and to identify optical/mid-infrared counterparts vary. It is difficult to disentangle the different approaches, but there is still very broad agreement in the conclusion that we see two different populations, one which, at lower redshifts, we find in VIKING counterparts, and the other, at higher redshifts, the fraction of which we can imply and which is consistent with at least some of the BLAST findings.\par
A similar picture emerges from the HerMES project \citep{oli10} so far. \citet{eal10} and \citet{elb10} use deep imaging in small areas ($<0.1 \sqdeg$) observed by SPIRE and with excellent multi-wavelength data available, as well as spectroscopic and photometric redshifts. Both groups use a 250 $\mic$ selected sample and assume a 24 $\mic$ detection. With a high fraction of spectroscopic redshifts ($>65\%$), \citet{elb10} find $35-40\%$ of their sources lie at $z>1$ (deduced from their fig. 2), consistent with our findings, whereas \citet{eal10} discover close to $50\%$ in this redshift range. 
\subsection{Towards more complete identifications} 

So far we have estimated that 73\% of SPIRE sources have
 counterparts in VIKING, while 51\% have a reliable match;
 thus, the reliable sample comprises approx 51/73 = 70 percent
 of all SPIRE galaxies with both $f_{250} > 32 \, {\rm mJy}$ and 
 $\Ks < 19.2$.  
 Of the remaining 49\% of SPIRE sources, 14\% are undetected in VIKING and 35\% have one or more low-reliability  match(es); overall we expect around half of these to be 
 genuine matches.  
 To make this decisive, we would need follow-up observations such as radio\footnote{Better positions, and greater efficiency of IDs will be possible with
the ASKAP radio survey EMU \citep{nor11} which will have 10"
angular resolution and cover a redshift range quite similar to
H-ATLAS.} or ALMA (Atacama Large Millimeter/submillimetre Array, \citealp{wot09}) imaging 
 giving a sub-arcsec position, or possibly
 optical/NIR spectroscopy of VIKING candidates (if we can 
  identify SPIRE sources via unusually strong emission lines). 
 This would normally be ``decisive'' in that a sub-arcsec radio/submm 
 position  would either match a VIKING galaxy to very high reliability,
  or if not it would prove the source is fainter than the VIKING limit. 
  
 For non-reliable matches, the total reliability $\sum R_i$ for a given 
  source is a good estimate of the probability that the SPIRE
 source has a real counterpart in VIKING, i.e. the probability
  that a follow-up will actually find a good match; therefore in 
 a follow-up search we should target the non-reliable matches in
 descending order of total probability. 
 
 Assuming a lower limit of the total probability of 50\% (70\%) would result in an additional 2380 (1856) identifications for 2967 (2101) observing targets. We would then obtain
 a sample which is 85\% (82\%) complete to $f_{250} > 32 \, {\rm mJy}$ and 
 $\Ks < 19.2$. If we would use all non-reliable positions as targets, regardless of the total reliability, we  would only be able to reach a completeness of 89\%. This effect can also be seen in \citet{smi10} where the sum of the reliabilities to all possible counterparts would result in a 44\% identification rate, lower than expected from their value of $Q_0=0.59$. This shows that our (and their) reliabilities might be underestimated; this is more evidence for a likely non-Gaussian positional error distribution which will be addressed in future publications of the catalogue.
 
Of additional consideration is our candidate list for multiple identifications, see section 5.3. They display a total reliability of 80\% by definition and would hence be included in a possible target list. It would be interesting to see how many of those could be confirmed as true multiple identifications.

\begin{figure}
\centering
\includegraphics[width=0.5\textwidth]{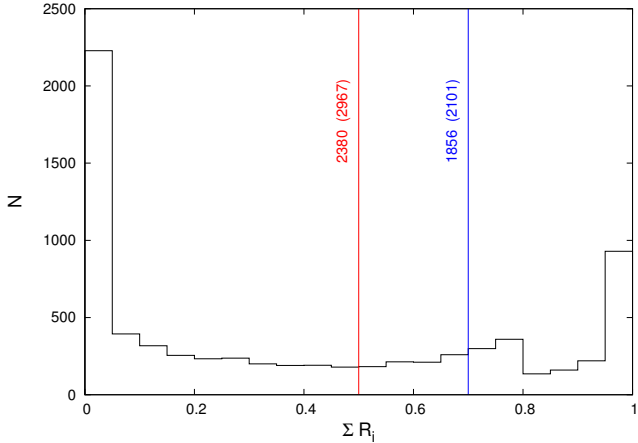}
\caption{--- Total reliability distribution of the non-reliable SPIRE sources. For the 7391 sources without a reliable counterpart, we sum the reliabilities of all possible VIKING counterparts per source. This total reliability is an estimate of the probability that the SPIRE source has a real counterpart in VIKING. If we include all SPIRE positions with total reliabilities greater than 50\% for a follow-up, we will have to observe 2967 positions and expect to gain 2380 additional reliable VIKING counterparts (red line and text). The blue line and text show the equivalent for a 70\% total reliability.} 
 \label{figTotRelDistr}
\end{figure}

Fig.~\ref{figTotRelDistr} shows the distribution
  of the total reliabilities for the non-reliable SPIRE positions together with the number of additional
   identifications we would expect if we followed up all SPIRE positions down to 50 or 70 percent.\par 
It is also useful to compare to the results of \citet{dun10}; as described above,
   they identified a much smaller sample of 20 BLAST $250 \mic$
   sources, but benefiting from the very deep multiwavelength data in
   GOODS-S.
   In their sample, all candidate identifications with $z < 1.2$
   are brighter than $\Ks < 19.6$, while all at $z > 1.2$ are fainter;
   this suggests that the 3011 sources without a VIKING counterpart have
   a high probability of being at $z > 1.2$ , and the same
   applies for the 3194 sources with low-reliability matches
   $\sum R < 0.2$.\par
   Since the flux ratio $f_{870}/f_{250}$ strongly increases with redshift
   for typical SMGs,
   the non-identified sources are therefore good targets for ALMA
   $870\mic$ follow-up snapshots; this could give a relatively
   efficient method for selecting luminous high-z SMGs.
   
  \section{Comparison to optical identifications \label{secComp}} 
  \subsection{Reliable counterparts \label{secCompRel}} 
  In this section we compare our results
with the optical identifications supplied with the H-ATLAS G09 Phase 1 source
catalogue. This used a similar likelihood ratio method with an
$r$-band selected sample down to $r=22.4$, as explained for the SDP data in \citet{smi10}. The VIKING $\Ks$-band should be
better placed than the optical $r$-band in identifying counterparts to the
SPIRE sources. As discussed in the introduction, at higher redshifts ($z \ge 0.5$), the $\Ks$ band is detecting
 flux from the near-infrared restframe, while the $r$ band is restframe blue/UV; 
 thus $\Ks$ is much better able to detect dusty galaxies. 
 We therefore expect a higher number of reliable identifications
 from matching with VIKING than with SDSS. To be able to compare our results,
we cross-match our VIKING candidate matches with the SDSS database (DR7) within 2''
and choose the nearest (primary) object.\par
\begin{table*}
\label{tabComp}
 \centering
 \caption{--- Comparison of the reliable counterparts to the SPIRE sources using optical r-band and near-infrared Ks-band matching. In the area corresponding to the VIKING preliminary source catalogue, we match reliably 51.3\% of all SPIRE sources. This is a significant increase in the identification rate compared to the 39.0\% of sources that are matched reliably to a SDSS object.}
  \begin{tabular}{lrrrrrrr}
  \hline
   & & & & N(reliable) & with SDSS &\multicolumn{2}{c}{N(reliable) not reliable in}\\
   band     &  N(SPIRE)  &  N(matches) & N(reliable)  &  in VIKING area & in VIKING area & $\Ks$  &  r    \\
 \hline
 $\Ks$ & 22,000 & 35,800  & 11,294 (51.3\%)& 11,294 (51.3\%)& 8,750 & - & 3,732\\
 r & 26,369 & 36,839 & 9,623 (36.5\%) & 8,587 (39.0\%)& 8,587 & 1,024 & - \\
 \hline
\end{tabular}
\end{table*}
We concentrate on the reliable counterparts 
 of both surveys. The Phase 1 catalogue lists 9623 reliable optical
counterparts ($36.5\%$) to 26,369 5$\sigma$ SPIRE sources in the G09 field; 
 of the reliable counterparts,  
 there are 8587 ($39.0\%$) in the VIKING observed area. 
 In comparison, we are able to match 11,294 SPIRE sources ($51.3\%$) reliably to VIKING $\Ks$ objects. \par
\begin{figure} \centering
\includegraphics[width=0.5\textwidth]{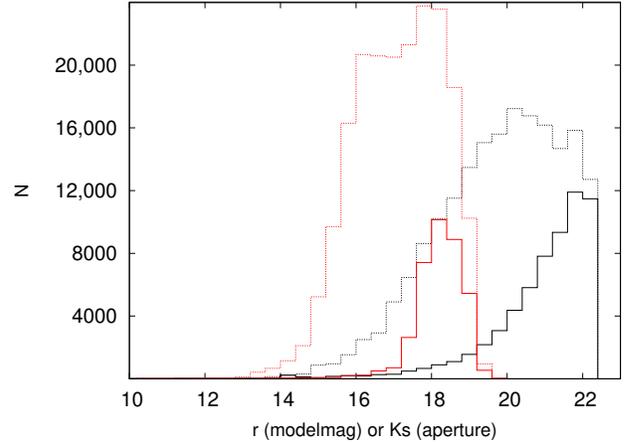}
\caption{--- The magnitude distributions of reliable $\Ks$ counterparts to SPIRE sources without
reliable optical counterparts. The black histograms show $r$ (modelmag, AB) distributions, the red histograms show
$\Ks$ (aperture, Vega) magnitude distributions. The dashed histograms represent
all reliable counterparts. The black solid line shows the r magnitude
distribution of the non-reliable optical matches 
to SPIRE sources with reliable $\Ks$
counterpart. The red solid line shows the distribution of $\Ks$ magnitudes
of optical drop-outs with reliable $\Ks$ counterpart.} \label{figCompMags}
\end{figure}

 We find a reliable $\Ks$ counterpart for 3732 SPIRE
sources without a reliable optical counterpart. Of the 3732 positions,
1717 are blank in SDSS ($\sim 21\%$ of all SDSS blank fields), 
 i.e. they are too faint to be detected in
 SDSS. Fig.~\ref{figCompMags} shows the $\Ks$ magnitude distribution of the
counterparts to the 1717 SPIRE positions that are optical dropouts (red
solid line). Unsurprisingly, the magnitudes are rather faint, 
 with a median of $\tilde{\Ks} = 18.26$, compared to
the magnitudes of all reliable $\Ks$ counterparts with a median of 
 $\tilde{\Ks} =17.07$.\par
 The remaining 2015 SPIRE  positions have optical counterparts, but their reliabilities lie below the
threshold of $R>0.8$. Fig.~\ref{figCompMags} (black solid line) shows the
$r$ modelmag distribution of the 3085 candidate matches to those 2015 SPIRE sources
from which we can see that they belong mostly to the faint end of the
overall magnitude distribution. \par
Conversely, there are 1,024 sources with optical reliable counterparts for which we did not find a reliable $\Ks$ counterpart. 
 Only 121 of those positions are blank in $\Ks$, mainly due
  to quality issues like saturation or bad pixels; the remaining
  903 sources share 2261 VIKING candidates, of which 706 have reliabilities with $0.5<R<0.8$.
 Comparing to our multiple candidates, see section~\ref{secResultsMultiples},
 we find that 590 of our 903 sources are indeed included in our candidate 
list of 1444 sources. 
 We also find 14 sources that have confirmed multiple counterparts 
 (by spectroscopic redshift).  
 Over half of the reliable matches we miss when compared to 
 the optical identifications, could hence be genuine multiple counterparts.\par
\begin{figure}
\centering
\includegraphics[width=0.5\textwidth]{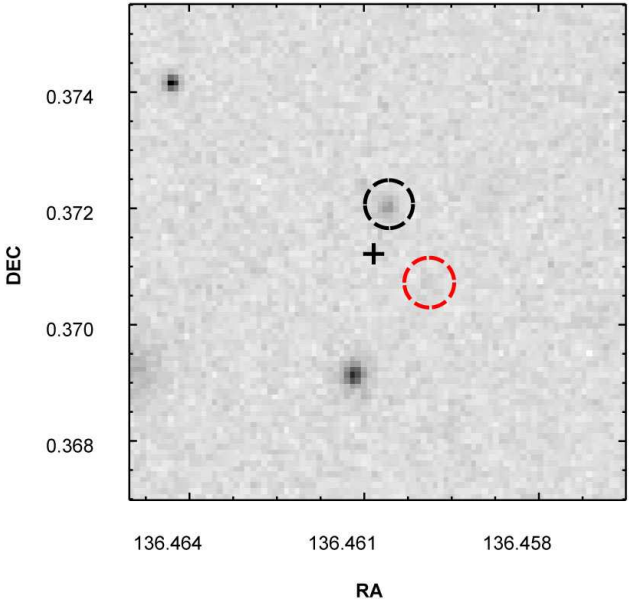}
\caption{--- VIKING $\Ks$ image of HATLAS J090550.5+002216, 30'' on the side. The black cross marks the SPIRE position. 
 Red circle: reliable SDSS counterpart ($R=0.85$), 
 black circle: reliable VIKING counterpart ($R=0.96$). 
 In the optical, the VIKING counterpart is detected with $r=23.16$, 
 fainter than the limit of $r=22.4$ used in the matching by \citet{smi10}. }
\label{figSDSSComp2}
\end{figure}

It is interesting to consider for how many SPIRE sources the
VIKING and SDSS matching disagree on reliable counterparts. 7563 SPIRE
sources ($\sim 88\%$ of the reliable optical matches in the VIKING area,
$\sim 67\%$ of the reliable VIKING matches) are matched reliably in both
surveys. Here, 7404 are matched to the same object. This leaves only
159 SPIRE sources (2.1\% of matches) 
 where the identification disagrees. Some of those are
deblending issues; often though we find that the reliable 
 optical counterpart is too faint in the
$\Ks$ band and/or the VIKING counterpart is too faint or not 
 detected in the $r$-band, resulting in different identifications. 
 Fig.~\ref{figSDSSComp2} shows an example,
HATLAS J090550.5+002216.

\subsection{Stellar counterparts \label{secCompStars}}
 Of our 12 reliable stellar matches, 7 have reliable SDSS counterparts.
  Here, 2 have been classified as galaxies in the Phase 1 catalogue, HATLAS J091233.9-004549 and HATLAS J085353.2+001648 (10930 and 21662 in Fig.~\ref{figStars}). Both are clearly stellar, showing diffraction spikes in both SDSS and VIKING images. The reliabilities of the stellar matches to the remaining 5 sources are high with $0.43<R<0.78$.\par 
Conversely, in the optical catalogue there are 21 reliable stellar matches: 
 of these, 19 are in the VIKING area, of which 5 are matched reliably to
  a VIKING star, 7 stellar objects are not included in our VIKING sample due to
  saturation in $\Ks$, and 
7 more are matched, but do not reach $R>0.8$ 
 (but have reliabilities $>0.4$).\par

\subsection{Photometric redshift comparison}
\begin{figure}
\centering
\includegraphics[width=0.5\textwidth]{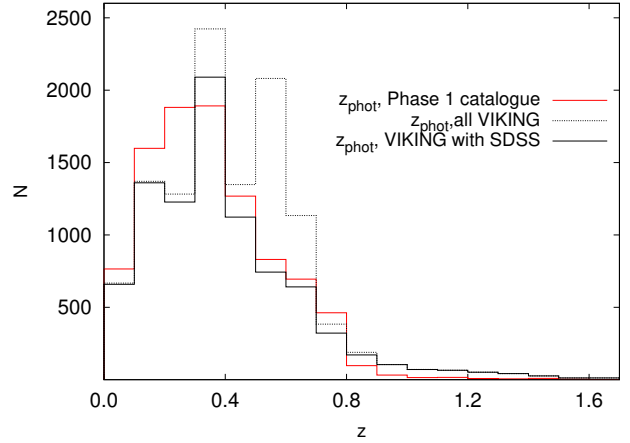}
\caption{--- Comparison of the photometric redshift distributions obtained in this work (VIKING) and supplied with the Phase 1 catalogue. The red line shows the latter, 9583 photometric redshifts with a median of $\tilde{z}=0.326$. The black dashed line shows the former, 11,294 photometric redshifts with a median of $\tilde{z}=0.396$. The black line represents photometric redshifts of the 8748 VIKING objects with SDSS counterparts.}
\label{figCompRed}
\end{figure}

Due to the facts that we match with SDSS to obtain our photometry and that ANNz is used in both cases, it is no surprise that the photometric redshift distributions of the reliable counterparts are broadly similar, see Fig.~\ref{figCompRed}. We obtain a slightly higher median redshift of $\tilde{z}=0.396$ compared to the median of $\tilde{z}=0.326$ of the photometric redshifts supplied with the Phase 1 catalogue, partly due to a higher number of redshifts $z_{phot}>1$. Of the 309 SPIRE sources with VIKING reliable counterparts and $z_{phot}>1$, 76 are reliable, have photometric redshifts and are matched to the same object in the optical catalogue. The photometric redshifts of those 76 objects differ by an average of 0.46. This large difference could be explained by incomplete photometric information from UKIDSS LAS used to compile the photometric redshifts in the optical catalogue. Indeed, nearly half (35) of the 76 objects have only 1 or 2 bands in the near-infrared available from LAS. This shows the advantage of the deeper VIKING data. Much better results should be achieved once we have optical photometry from the VST KIDS survey to combine with VIKING.

\section{Conclusions\label{secCon}} 
 We
 have matched the 22,000 SPIRE G09 sources that fall within the VIKING observed
area to a catalogue of near-infrared objects from VIKING: 
 the VIKING sample contains 1,376,606 objects,  
  classified into 847,530 galaxies and 529,076 stars
  according to shape and colour parameters 
 using a modified version of the method of \citet{bal10}. 
 We found statistically, using blank-field comparisons, 
  that $73 \pm 2\%$ of SPIRE sources should have one or more counterpart 
  detections in VIKING. 
 With a search radius of $10''$ we found 35,800 candidate matches:  
  applying a likelihood ratio method to calculate the
 probability of each candidate to be the true counterpart of the SPIRE source,
  we find matches to 11,294 sources ($51.3\%$) with a probability $>80\%$,
  or reliability $R>0.8$. The false identification rate is estimated to be
 4.2\% and the probability of mis-identifying 
  a true counterpart is $\sim 5\%$. 
 \par
 Of the reliable counterparts, 3147 (27.9\%) have spectroscopic 
 redshifts from either the
 GAMA or the SDSS redshift surveys. We calculate photometric redshifts for
the remaining possible matches, using a sample of 32,465 spectroscopic redshifts as a
training set. 
The errors in the redshift estimation are investigated using the
existing 3147 spectroscopic redshifts. 
We find a scatter of $\sigma = 0.0353$
in the difference $|z_{spec}-z_{phot}|$ which is comparable with $\sigma =
0.037$ found by \citet{smi10} when calculating photometric redshifts from
SDSS/UKIDSS LAS photometry for the HATLAS SDP field. For $z\gtrsim 0.3$ we find that photometric redshifts are systematically underestimated with a bias of $\sim 0.09$. \par
 Comparing our results with that from the $r$-band matching
to SDSS objects supplied with the SPIRE catalogue, we report a $\gtrsim 12\%$
increase in the reliable identification rate.
 We find that we agree on reliable counterparts for
  $\sim 88\%$ of the reliable optical matches to sources within the
  VIKING area of the G09 field.\par
     The identifications here provide a useful potential preselection for
   follow-up studies: the moderate-reliability matches could mostly
   be confirmed or rejected using optical multi-object spectroscopy, giving
   a mostly complete subsample for sources at $z < 1$;
    while the $\sim 28\%$ of sources with no match or low-reliability match(es)
   have a high probability of being at $z > 1$, and form a
   large sample of interesting targets for ALMA $870\mic$ snapshots.\par

A future SPIRE source catalogue
will include the VIKING ZYJH$\Ks$ photometry (aperture magnitudes)
for all candidate matches, but we stress that only matches with $R>0.8$ should be
regarded as reliable counterparts and used for science application.\par

\section{Acknowledgements\label{secAck}}
We thank the anonymous referee for useful comments.\par
The \textit{Herschel}-ATLAS is a project with \textit{Herschel}, which is an ESA space observatory with science instruments provided by European-led Principal Investigator consortia and with important participation from NASA. The H-ATLAS website is http://www.h-atlas.org/ U.S. participants in \textit{Herschel}–ATLAS acknowledge support provided by NASA through a contract
issued from JPL. \par
GAMA is a joint European-Australasian
project based around a spectroscopic campaign using the Anglo-Australian Telescope. 
The GAMA input catalogue is based on data
taken from the Sloan Digital Sky Survey and the UKIRT Infrared
Deep Sky Survey. Complementary imaging of the GAMA regions
is being obtained by a number of independent survey programs
including GALEXMIS, VST KIDS, VISTA VIKING,WISE,
Herschel-ATLAS, GMRT and ASKAP providing UV to radio coverage.
The GAMA website is: http://www.gama-survey.org/. This work used data from the UKIDSS DR5 and the SDSS
DR7. The UKIDSS project is defined in Lawrence et al. (2007) and
uses the UKIRTWide Field Camera (WFCAM; Casali et al. 2007).
Funding for the SDSS and SDSS-II has been provided by the Alfred
P. Sloan Foundation, the Participating Institutions, The National
Science Foundation, the U.S. Department of Energy, the National
Aeronautics and Space Administration, the Japanese Monbukagakusho,
the Max Planck Society and the Higher Education
Funding Council for England.\par 
The Italian group acknowledges partial
financial support from ASI contract I/009/10/0 'COFIS'. \par
This work used data from VISTA at the ESO Paranal Observatory,  
 programme 179.A-2004. 
VISTA is an ESO near-infrared survey telescope in Chile, 
 conceived and developed by a consortium of universities in the 
 United Kingdom, led by Queen Mary University, London.

\bsp
\onecolumn
\appendfigs

\suppressfloats[t]
\begin{figure}
\centering
\subfloat[][\textbf{J084347.6+005019:}\hspace{0.1in}$z_{\rm spec}=$]{
\includegraphics[width=0.2\textwidth,height=0.2\textwidth]{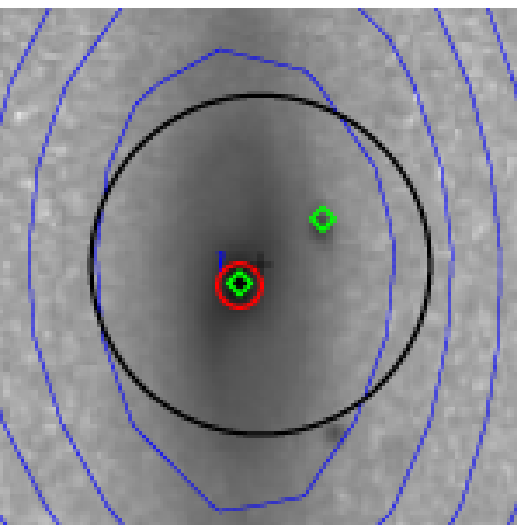}
\label{fig:subfig2a}}
\subfloat[][0.029]{
\includegraphics[width=0.2\textwidth,height=0.2\textwidth]{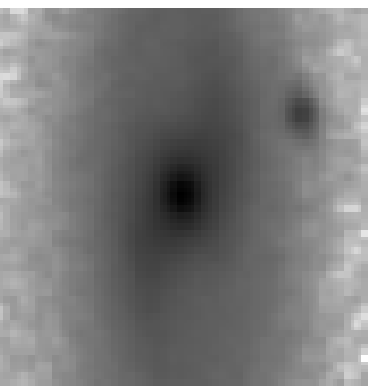}
\label{fig:subfig2b}}
\subfloat[][]{
\includegraphics[width=0.2\textwidth,height=0.2\textwidth]{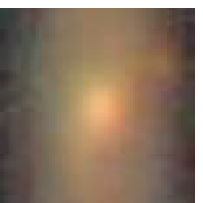}
\label{fig:subfig2c}}\\
\subfloat[][\textbf{J092344.5-003104:}\hspace{0.1in}$z_{\rm spec}=$]{
\includegraphics[width=0.2\textwidth,height=0.2\textwidth]{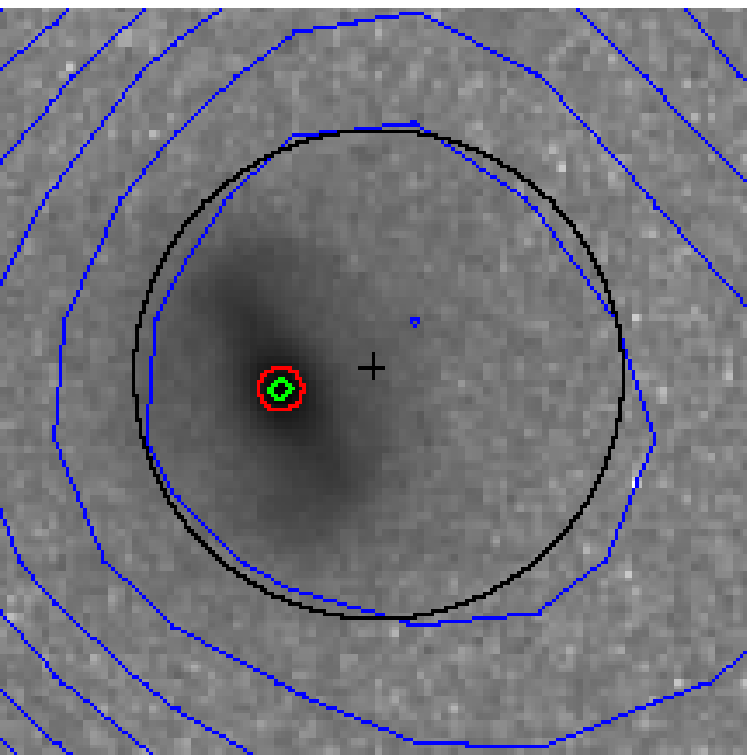}
\label{fig:subfig5a}}
\subfloat[][0.056]{
\includegraphics[width=0.2\textwidth,height=0.2\textwidth]{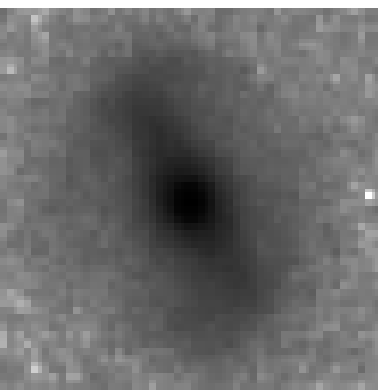}
\label{fig:subfig5b}}
\subfloat[][]{
\includegraphics[width=0.2\textwidth,height=0.2\textwidth]{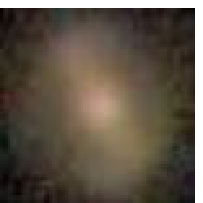}
\label{fig:subfig5c}}\\
\subfloat[][\textbf{J085116.1-001410:}\hspace{0.1in}$z_{\rm spec}=$]{
\includegraphics[width=0.2\textwidth,height=0.2\textwidth]{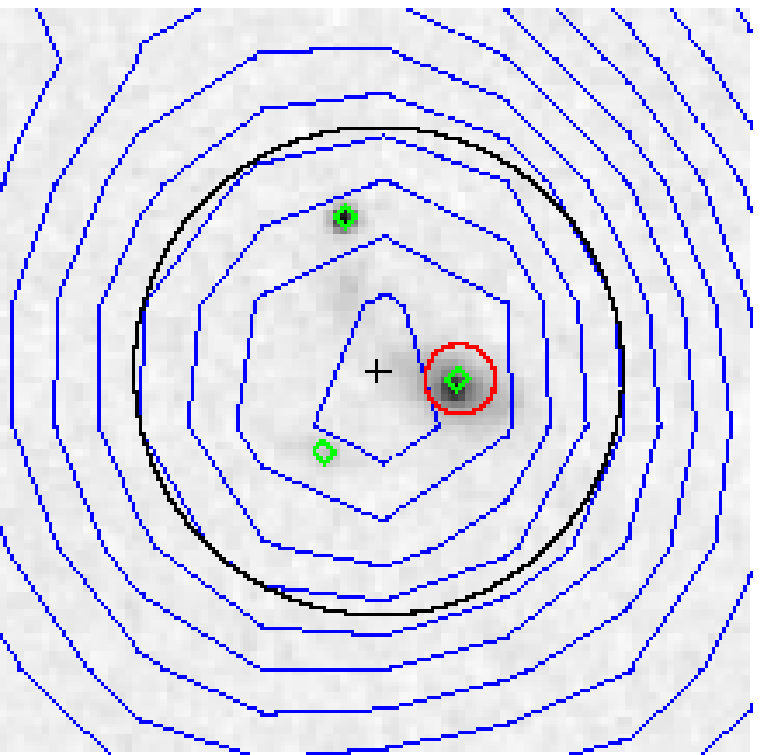}
\label{fig:subfig7a}}
\subfloat[][0.268]{
\includegraphics[width=0.2\textwidth,height=0.2\textwidth]{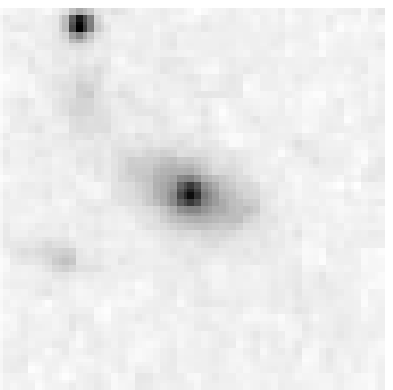}
\label{fig:subfig7b}}
\subfloat[][]{
\includegraphics[width=0.2\textwidth,height=0.2\textwidth]{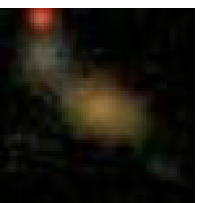}
\label{fig:subfig7c}}\\
\subfloat[][\textbf{J084706.4+021212:}\hspace{0.1in}$z_{\rm spec}=$]{
\includegraphics[width=0.2\textwidth]{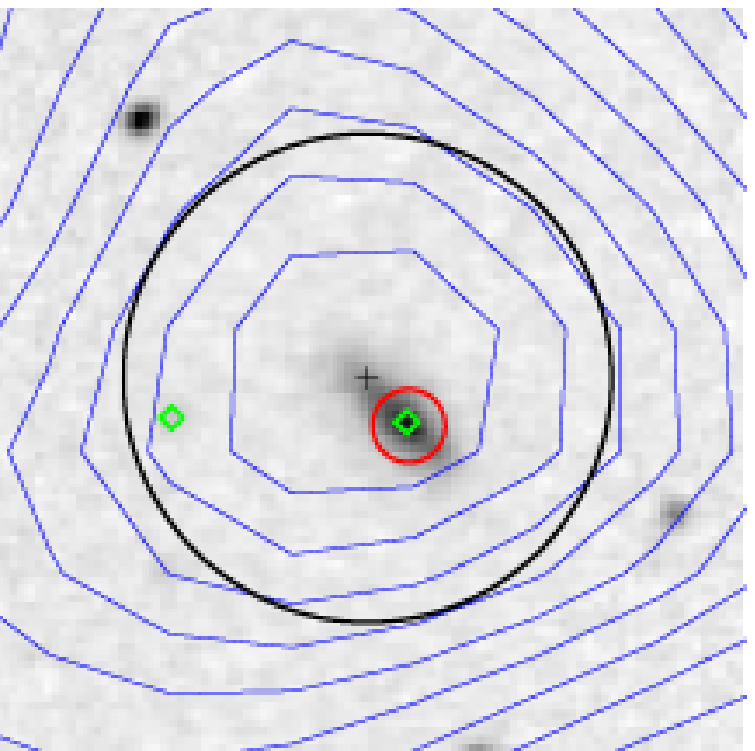}
\label{fig:subfig8a}}
\subfloat[][0.074]{
\includegraphics[width=0.2\textwidth,height=0.2\textwidth]{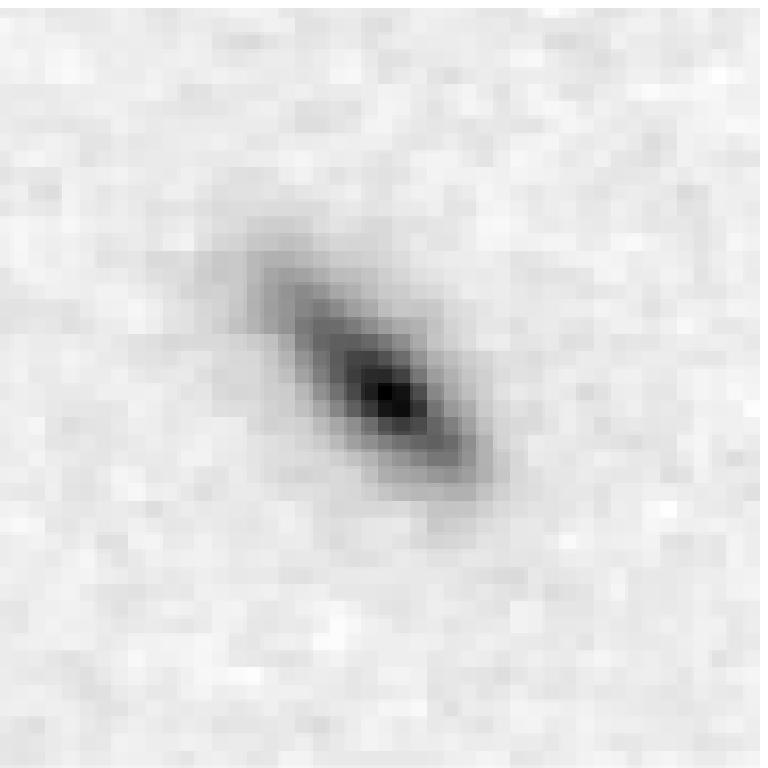}
\label{fig:subfig8b}}
\subfloat[][]{
\includegraphics[width=0.2\textwidth,height=0.2\textwidth]{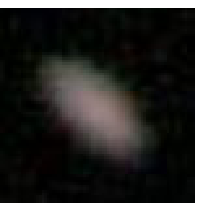}
\label{fig:subfig8c}}\\
\caption{VIKING and SDSS cutouts of a random selection of SPIRE positions. The image on the \textbf{left} shows a 30'' VIKING cutout; the possible counterparts are indicated with a green diamond, the reliable counterpart is encircled in red. The black (or white) cross shows the SPIRE position and the black circle represents the search radius of 10''. The \textbf{middle} image shows a 15'' VIKING cutout centered on the position of the reliable counterpart. The image on the \textbf{right} shows a 15'' SDSS cutout centered on the position of the reliable VIKING counterpart. On all images, North is up and East is left.}
\end{figure}
\begin{figure}
\ContinuedFloat
\centering
\subfloat[][\textbf{J091235.5+020450:}\hspace{0.1in}$z_{\rm phot}=$]{
\includegraphics[width=0.2\textwidth,height=0.2\textwidth]{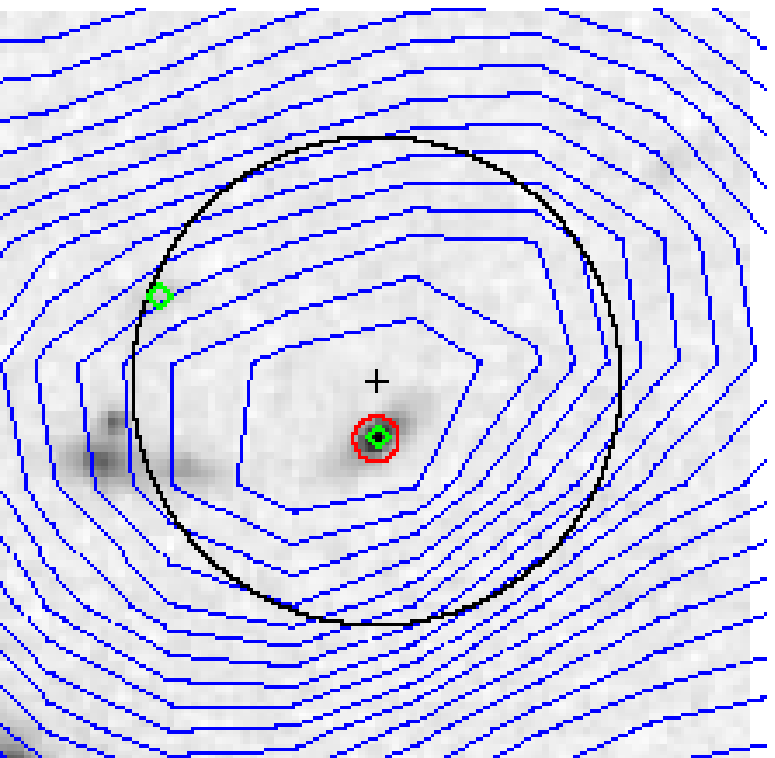}
\label{fig:subfig9a}}
\subfloat[][0.346]{
\includegraphics[width=0.2\textwidth,height=0.2\textwidth]{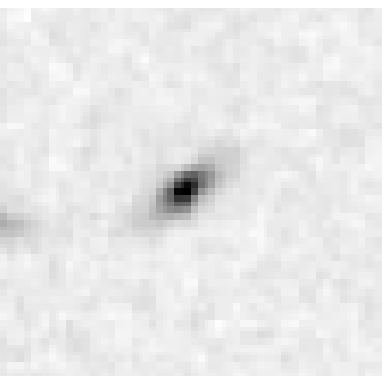}
\label{fig:subfig9b}}
\subfloat[][]{
\includegraphics[width=0.2\textwidth,height=0.2\textwidth]{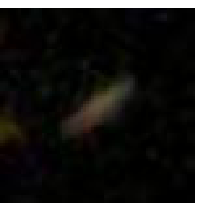}
\label{fig:subfig9c}}\\
\subfloat[][\textbf{J084750.0-002242:}\hspace{0.1in}$z_{\rm phot}=$]{
\includegraphics[width=0.2\textwidth,height=0.2\textwidth]{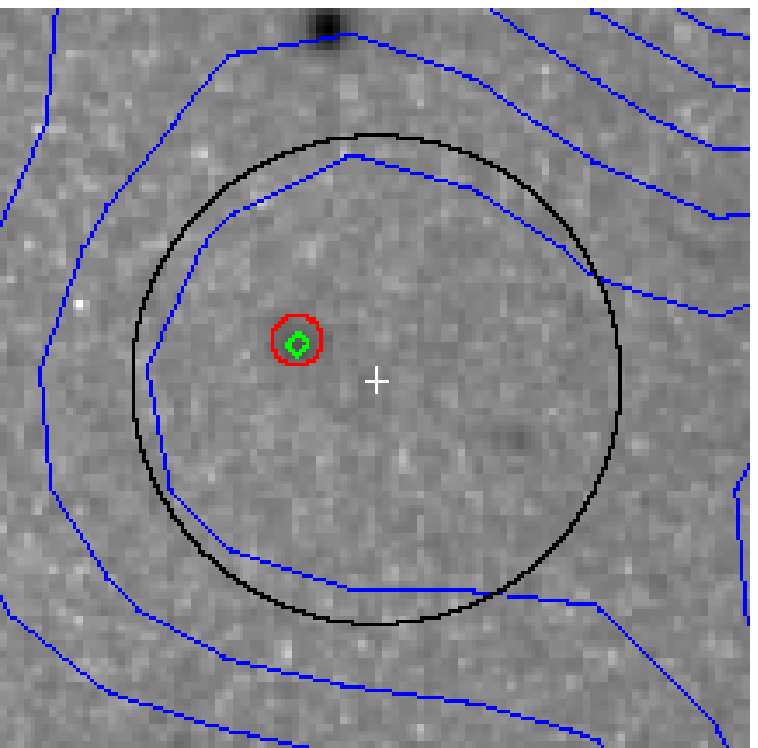}
\label{fig:subfig12a}}
\subfloat[][1.131]{
\includegraphics[width=0.2\textwidth,height=0.2\textwidth]{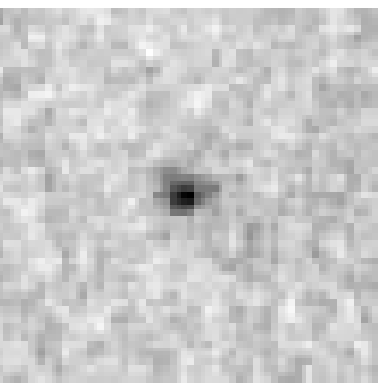}
\label{fig:subfig12b}}
\subfloat[][]{
\includegraphics[width=0.2\textwidth,height=0.2\textwidth]{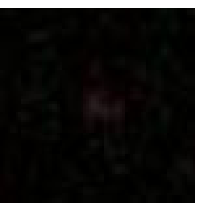}
\label{fig:subfig12c}}\\
\subfloat[][\textbf{J092124.4+011600:}\hspace{0.1in}$z_{\rm phot}=$]{
\includegraphics[width=0.2\textwidth,height=0.2\textwidth]{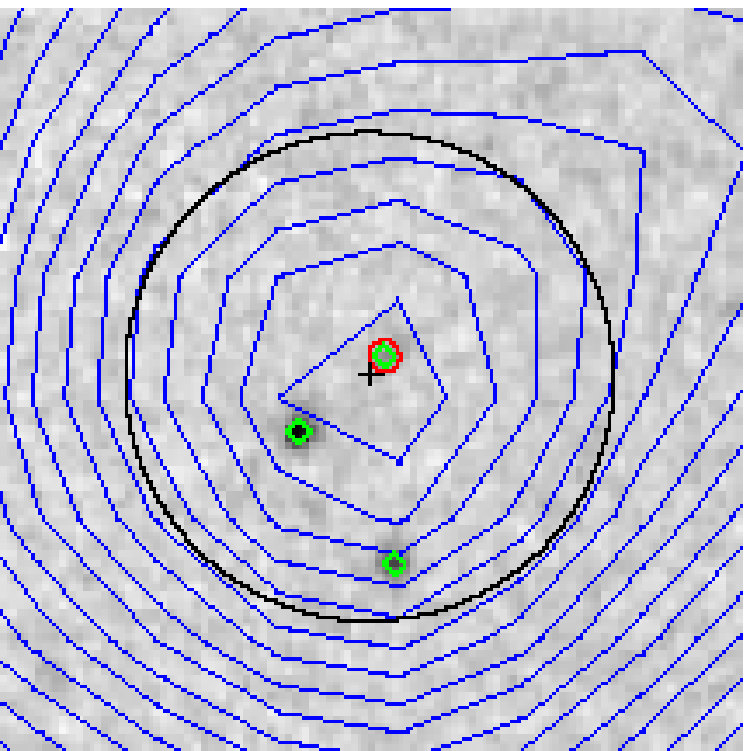}
\label{fig:subfig14a}}
\subfloat[][0.565]{
\includegraphics[width=0.2\textwidth,height=0.2\textwidth]{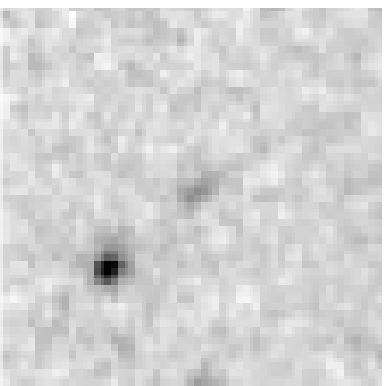}
\label{fig:subfig14b}}
\subfloat[][]{
\includegraphics[width=0.2\textwidth,height=0.2\textwidth]{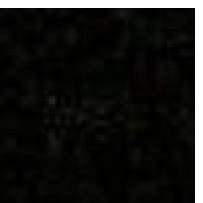}
\label{fig:subfig14c}}\\
\subfloat[][\textbf{J083848.1+014536:}\hspace{0.1in}$z_{\rm phot}=$]{
\includegraphics[width=0.2\textwidth,height=0.2\textwidth]{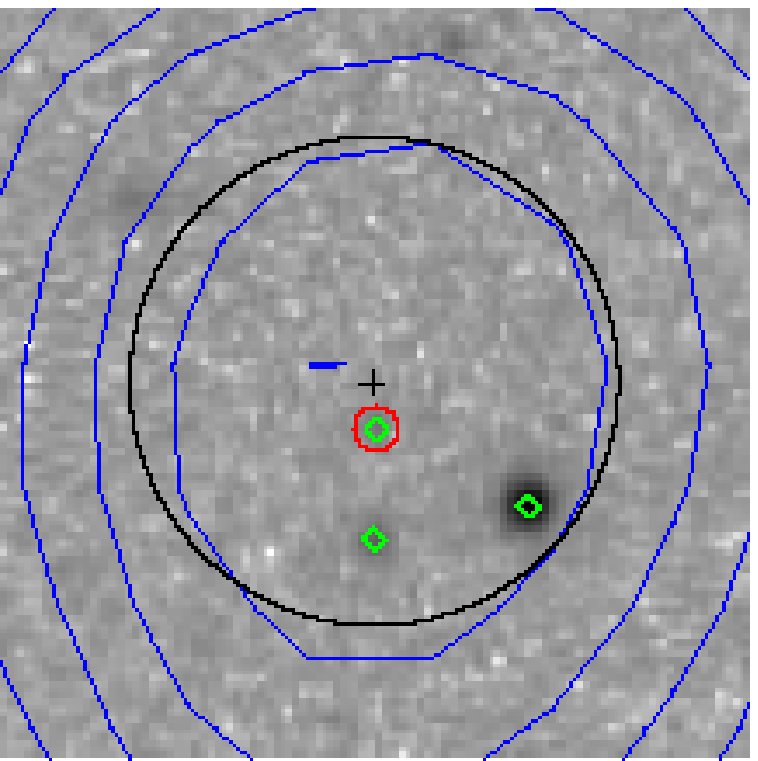}
\label{fig:subfig15a}}
\subfloat[][0.546]{
\includegraphics[width=0.2\textwidth,height=0.2\textwidth]{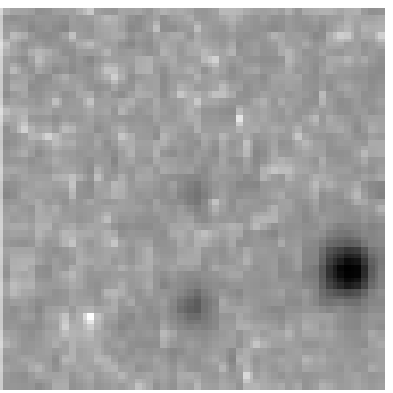}
\label{fig:subfig15b}}
\subfloat[][]{
\includegraphics[width=0.2\textwidth,height=0.2\textwidth]{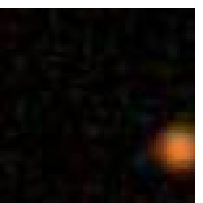}
\label{fig:subfig15c}}\\
\subfloat[][\textbf{J091858.3+013454:}\hspace{0.1in}$z_{\rm phot}=$]{
\includegraphics[width=0.2\textwidth,height=0.2\textwidth]{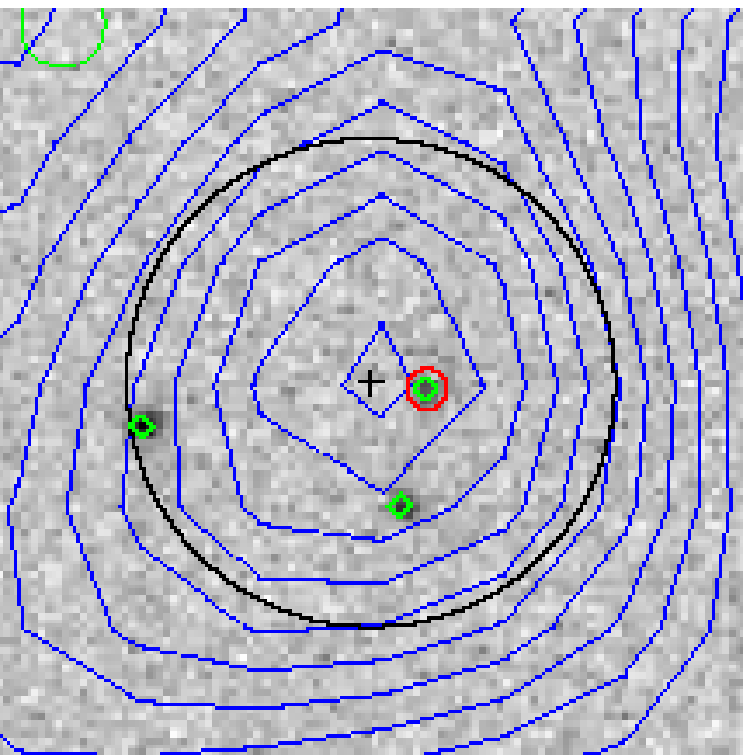}
\label{fig:subfig17a}}
\subfloat[][0.814]{
\includegraphics[width=0.2\textwidth,height=0.2\textwidth]{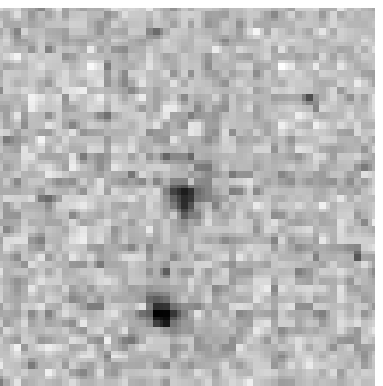}
\label{fig:subfig17b}}
\subfloat[][]{
\includegraphics[width=0.2\textwidth,height=0.2\textwidth]{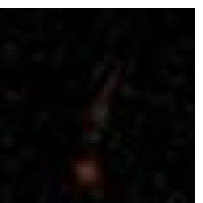}
\label{fig:subfig17c}}\\
\caption{continued}
\end{figure}

\label{lastpage}


\begin{thebibliography}{99}
\bibitem[\protect\citeauthoryear{Amblard et al.}{2010}]{amb10} Amblard A. et al., 2010,
A\&A, 518, L9
\bibitem[\protect\citeauthoryear{Baldry et al.}{2010}]{bal10} Baldry I.K. et al., 2010,
MNRAS, 404, 86
\bibitem[\protect\citeauthoryear{Bell et al.}{2003}]{bel03} Bell E.F., McIntosh D.H., Katz N., Weinberg M., 2003, ApJS, 149, 289
\bibitem[\protect\citeauthoryear{Bell et al.}{2006}]{bel06} Bell E.F., Phleps S., Somerville R.S., Wolf C., Borch A., Meisenheimer K., 2006, ApJ, 652, 270
\bibitem[\protect\citeauthoryear{Biggs et al.}{2011}]{big11} Biggs A.D. et al., 2011,
MNRAS, 413, 2314
\bibitem[\protect\citeauthoryear{Blain \& Longair}{1993}]{blr93} Blain A.W., Longair M.S., 1993,
MNRAS, 264, 509
\bibitem[\protect\citeauthoryear{Bond et al.}{submitted to ApJL}]{bonprep} Bond N. et al., submitted to ApJL
\bibitem[\protect\citeauthoryear{Bourne et al.}{2011}]{bou11} Bourne N. et al., 2011, MNRAS, 410, 1155 \bibitem[\protect\citeauthoryear{Chapin et al.}{2011}]{cha11} Chapin E.L. et al., 2011,
 MNRAS, 411, 505
\bibitem[\protect\citeauthoryear{Ciliegi et al.}{2003}]{cil03} Ciliegi P., Zamorani G., Hasinger G., Lehmann I., Szokoly G., Wilson G., 2003, A\&A, 398, 901 
\bibitem[\protect\citeauthoryear{Ciliegi et al.}{2005}]{cil05} Ciliegi P. et al., 2005,
 A\&A, 441, 879 
\bibitem[\protect\citeauthoryear{Cole et al.}{2001}]{col01} Cole S. et al., 2001, MNRAS, 326, 255
\bibitem[\protect\citeauthoryear{Collister \& Lahav}{2004}]{col04} Collister A., Lahav O., 2004,
PASP, 116, 818, 345 
\bibitem[\protect\citeauthoryear{DeBoer et al.}{2009}]{deb09} DeBoer D.R., 2009, Proc. IEEE, 97, 1507 
\bibitem[\protect\citeauthoryear{de Ruiter, Arp \& Willis}{1977}]{deR77} de Ruiter H.R., Arp H. C., Willis
A. G., 1977, A\&AS, 28, 211
\bibitem[\protect\citeauthoryear{Dariush et al.}{2011}]{dar11} Dariush A. et al., 2011, MNRAS, 418, 64
\bibitem[\protect\citeauthoryear{Davis et al.}{2003}]{dav03} Davis M. et al., 2003, Proc. SPIE, 4834, 161
\bibitem[\protect\citeauthoryear{Devlin et al.}{2009}]{dev09} Devlin M. et al., 2009, Nature, 458, 737
\bibitem[\protect\citeauthoryear{Driver et al.}{2011}]{dri11} Driver S.P. et al., 2011, MNRAS, 412,765
\bibitem[\protect\citeauthoryear{Dunlop et al.}{2010}]{dun10} Dunlop J.S. et al., 2010, MNRAS, 408, 2022
\bibitem[\protect\citeauthoryear{Dye et al.}{2009}]{dye09} Dye S. et al., 2009, ApJ, 703, 285
\bibitem[\protect\citeauthoryear{Dye et al.}{2010}]{dye10} Dye S. et al., 2010, A\&A, 518, L10
\bibitem[\protect\citeauthoryear{Eales et al.}{2010}]{eal10} Eales S.A. et al., 2010, PASP, 122, 499
\bibitem[\protect\citeauthoryear{Elbaz et al.}{2010}]{elb10} Elbaz D. et al., 2010, A\&A, 518, L29
\bibitem[\protect\citeauthoryear{Emerson et al.}{2004}]{eme04} Emerson J.P. et al., 2004, Proc. SPIE, 5493, 401
\bibitem[\protect\citeauthoryear{Emerson \& Sutherland}{2010}]{eme10} Emerson J.P., Sutherland W.J., 2010, Proc. SPIE, 7733, 4
\bibitem[\protect\citeauthoryear{Findlay et al.}{2012}]{fin12} Findlay, J. R., Sutherland, W. J., Venemans, B. P., Reyl\'{e}, C., Robin, A. C., Bonfield, D. G., Bruce, V. A., Jarvis, M. J., MNRAS, 419, 3354
\bibitem[\protect\citeauthoryear{Fukugita et al.}{1996}]{fuk96} Fukugita M., Ichikawa, T., Gunn, J.E., Doi, M., Shimasaku, K., Schneider, D.P., 1996, AJ, 111, 1748
\bibitem[\protect\citeauthoryear{Gonz\'{a}les-Nuevo et al.}{2010}]{gon10} Gonz\'{a}les-Nuevo J. et al., 2010, A\&A, 518, L38
\bibitem[\protect\citeauthoryear{Griffin et al.}{2010}]{gri10} Griffin M.J. et al., 2010, A\&A, 518, L3
\bibitem[\protect\citeauthoryear{Hainline et al.}{2010}]{hai10} Hainline L.J. et al., 2010, AAS 21532601, 42, 422
\bibitem[\protect\citeauthoryear{Hambly et al.}{2008}]{ham08} Hambly N. et al., 2008, MNRAS, 384, 637
\bibitem[\protect\citeauthoryear{Hardcastle et al.}{2010}]{har10} Hardcastle M.J. et al., 2010, MNRAS, 409, 122
\bibitem[\protect\citeauthoryear{Helou et al.}{1985}]{hel85} Helou G., Soifer B.T., Rowan-Robinson M., 1985, ApJ, 298, 7
\bibitem[\protect\citeauthoryear{Hewitt et al.}{2006}]{hew06} Hewitt P.C. et al., 2006, MNRAS, 367, 454
\bibitem[\protect\citeauthoryear{Hoyos et al.}{in prep.}]{hoyinPrep} Hoyos C. et al., in preparation
\bibitem[\protect\citeauthoryear{Ibar et al.}{2010}]{iba10} Ibar E. et al., 2010, MNRAS, 409, 38
\bibitem[\protect\citeauthoryear{Ivison et al.}{2007}]{ivi07} Ivison R.J. et al., 2007, MNRAS, 380, 199
\bibitem[\protect\citeauthoryear{Ivison et al.}{2010}]{ivi10} Ivison R.J. et al., 2010, A\&A, 518, L31
\bibitem[\protect\citeauthoryear{Jarvis et al.}{2010}]{jar10} Jarvis M. et al., 2010, MNRAS, 409, 92
\bibitem[\protect\citeauthoryear{Johnston et al.}{2008}]{joh08} Johnston S. et al., 2008, Experimental Astronomy, 22, Issue 3, 151
\bibitem[\protect\citeauthoryear{Kim et al.}{submitted to ApJ}]{kimprep} Kim S. et al., submitted to ApJ, arXiv:1112.3653
\bibitem[\protect\citeauthoryear{Lagache et al.}{2003}]{lag03} Lagache G., Dole H., Puget J.-L., 2003, MNRAS, 338, 555
\bibitem[\protect\citeauthoryear{Lapi et al.}{2011}]{lap11} Lapi A. et al., ApJ, 742, 24
\bibitem[\protect\citeauthoryear{Lawrence et al.}{2010}]{law07} Lawrence A. et al., 2007, MNRAS, 379, 1599
\bibitem[\protect\citeauthoryear{Lewis, Irwin \& Bunclark}{2010}]{lew10} Lewis J.R., Irwin M., Bunclark P., 2010, ASPC, 434, 91L
\bibitem[\protect\citeauthoryear{Lilly et al.}{2007}]{lil07} Lilly S.J., 2007, ApJS, 172, 70
\bibitem[\protect\citeauthoryear{Lotz}{2007}]{lot07} Lotz J.M., 2007, ASPC, 380, 467
\bibitem[\protect\citeauthoryear{Maddox et al.}{2010}]{mad10} Maddox S.J. et al., 2010, A\&A, 518, L11
\bibitem[\protect\citeauthoryear{Michalowski, Watson \& Hjorth}{2010}]{mic10} Michalowski M.J., Watson D. and Hjorth J. 2010, ApJ, 712, 942
\bibitem[\protect\citeauthoryear{Mortier et al.}{2005}]{mor05} Mortier A.M.J. et al., 2005, MNRAS, 363, 563
\bibitem[\protect\citeauthoryear{Negrello et al.}{2007}]{neg07} Negrello M. et al., 2007, MNRAS, 377, 1557
\bibitem[\protect\citeauthoryear{Negrello et al.}{2010}]{neg10} Negrello M. et al., 2010, Sci, 330, 800
\bibitem[\protect\citeauthoryear{Norris et al.}{2011}]{nor11} Norris R.P. et al., 2011, PASA, 28, 215
\bibitem[\protect\citeauthoryear{Oliver et al.}{2010}]{oli10} Oliver M. et al., 2010, A\&A 518, L21
\bibitem[\protect\citeauthoryear{Pascale et al.}{2010}]{pas10} Pascale E. et al., 2011, MNRAS, 415, 911
\bibitem[\protect\citeauthoryear{Pilbratt et al.}{2010}]{pil10} Pilbratt G.L. et al., 2010, A\&A, 518 L1
\bibitem[\protect\citeauthoryear{Poglitsch et al.}{2010}]{pog10} Poglitsch A. et al., 2010, A\&A, 518, L2
\bibitem[\protect\citeauthoryear{Prestage \& Peacock}{1983}]{pp83} Prestage R.M., Peacock J. A.,
1983, MNRAS, 204, 355
\bibitem[\protect\citeauthoryear{Rigby et al.}{2011}]{rig10} Rigby E.E. et al., 2011, MNRAS, 415, 3
\bibitem[\protect\citeauthoryear{Rodighiero et al.}{2010}]{rod10} Rodighiero G. et al., 2010,  A\&A 518, L25
\bibitem[\protect\citeauthoryear{Roseboom et al.}{2010}]{ros10} Roseboom I.G. et al., 2010, MNRAS, 409, 48
\bibitem[\protect\citeauthoryear{Ryan et al.}{2008}]{rya08} Ryan R.E., Cohen S.H., Windhorst R.A., Silk J., 2008, ApJ, 678, 751
\bibitem[\protect\citeauthoryear{Schilizzi, Dewdney \& Lazio}{2008}]{sch08} Schilizzi Richard T., Dewdney Peter E.F., Lazio T. Joseph W., 2008, Proc. SPIE, 7012, 52
\bibitem[\protect\citeauthoryear{Serjeant et al.}{2003}]{ser03} Serjeant S. et al., 2003, MNRAS, 334, 887
\bibitem[\protect\citeauthoryear{Smail, Ivison \& Blain}{1997}]{sma97} Smail I., Ivison R.J., Blain A.W., 1997, ApJ, 490, L5
\bibitem[\protect\citeauthoryear{Smith et al.}{2011a}]{smi10} Smith D.J.B. et al., 2011a, MNRAS, 416, 857
\bibitem[\protect\citeauthoryear{Smith et al.}{2011b}]{smiinPrep} Smith D.J.B. et al., 2011b, submitted to MNRAS
\bibitem[\protect\citeauthoryear{Sutherland \& Saunders}{1992}]{ss92} Sutherland W., Saunders W., 1992, MNRAS, 259, 413
\bibitem[\protect\citeauthoryear{Sutherland}{in prep.}]{sutprep} Sutherland W. et al., in preparation
\bibitem[\protect\citeauthoryear{Thompson et al.}{2010}]{tho10} Thompson M.A. et al., 2010, A\&A, 518, L134
\bibitem[\protect\citeauthoryear{V\'{e}ron-Cetty \& V\'{e}ron}{2010}]{ver10} V\'{e}ron-Cetty M.-P., V\'{e}ron P., 2010, A\&A, 518, A10
\bibitem[\protect\citeauthoryear{Wang \& Rowan-Robinson}{2009}]{wrr09} Wang L., Rowan-Robinson M., 2009, MNRAS, 398, 109
\bibitem[\protect\citeauthoryear{Wolstencroft et al.}{1986}]{wol86} Wolstencroft R.D., Savage A., Clowes R.G., MacGillivray H.T., Leggett S.K., Kalafi M., 1986, MNRAS, 223, 279
\bibitem[\protect\citeauthoryear{Wotten \& Thompson}{2009}]{wot09} Wotten A., Thompson A.R., 2009, Proc. IEEE, 97, 1463
\bibitem[\protect\citeauthoryear{Wright et al.}{2010}]{wri10} Wright E.L. et al., 2010, AJ, 140, 1868
\bibitem[\protect\citeauthoryear{York et al.}{2000}]{yor00} York D.G. et al., 2000, AJ, 120, 1579
\end{thebibliography}
\end{document}